\documentclass[journal,a4paper,doublecolumn]{IEEEtran}
\IEEEoverridecommandlockouts
\usepackage{amssymb}
\usepackage{amsfonts}
\usepackage{epstopdf}
\usepackage{graphicx}
\usepackage{stmaryrd}
\usepackage{amssymb,amsmath}
\usepackage{multirow,url}
\usepackage{color,cite}
\usepackage{hyperref}
\usepackage{comment}
\usepackage{subcaption}
\usepackage{slashbox} 

\ifodd 1
\newcommand{\com}[1]{\textbf{\color{red} (COMMENT: #1)}} 
\newcommand{\comg}[1]{\textbf{\color{green} (COMMENT: #1)}}
\newcommand{\response}[1]{\textbf{\color{magenta} (RESPONSE: #1)}} 
\else

\newcommand{\com}[1]{}
\newcommand{\comg}[1]{}
\newcommand{\response}[1]{}
\fi
\ifodd 0
\newcommand{\referred}[1]{\textcolor{red}{RefPaper: #1}} 
\else
\newcommand{\referred}[1]{}
\fi

\ifodd 0
\newcommand{\changeblue}[1]{\textcolor{blue}{Modified: #1}} 
\else
\newcommand{\changeblue}[1]{}
\fi

\begin{document}

\title{Network-Coded Multiple Access with High-order Modulations}


\author{Haoyuan~Pan,~\IEEEmembership{Student Member,~IEEE,}~Lu~Lu,~\IEEEmembership{Member,~IEEE,}
       and~Soung~Chang~Liew,~\IEEEmembership{Fellow,~IEEE}
\thanks{H. Pan and S. C. Liew are with the Department of Information Engineering, The Chinese University of Hong Kong, Hong Kong.
Email: \{ph014, soung\}@ie.cuhk.edu.hk. L. Lu is with the Institute of Network Coding, The Chinese University of Hong Kong, Hong Kong.
Email: lulu@ie.cuhk.edu.hk. 
}
}

\maketitle

\begin{abstract}
This paper presents the first network-coded multiple access (NCMA) system prototype operated on high-order modulations up to 16-QAM. NCMA jointly exploits physical-layer network coding (PNC) and multiuser decoding (MUD) to boost throughput of multipacket reception systems. Direct generalization of the existing NCMA decoding algorithm, originally designed for BPSK, to high-order modulations, will lead to huge performance degradation. The throughput degradation is caused by the relative phase offset between received signals from different nodes. To circumvent the phase offset problem, this paper investigates an NCMA system with multiple receive antennas at the access point (AP), referred to as \emph{MIMO-NCMA}. We put forth a low-complexity \emph{symbol-level} NCMA decoder that, together with MIMO, can substantially alleviate the performance degradation induced by relative phase offset. To demonstrate the feasibility and advantage of MIMO-NCMA for high-order modulations, we implemented our designs on software-defined radio. Our experimental results show that the throughput of QPSK MIMO-NCMA is double that of both BPSK NCMA and QPSK MUD at SNR=10dB. For higher SNRs at which 16-QAM can be supported, the throughput of MIMO-NCMA can be as high as 3.5 times that of BPSK NCMA. Overall, this paper provides an implementable framework for high-order modulated NCMA.
\end{abstract}



\IEEEpeerreviewmaketitle



\section{Introduction}

Multipacket reception is conventionally realized by multiuser decoding (MUD) techniques using orthogonal signaling \referred{Verdubook}\cite{Verdubook} (e.g., TDMA, CDMA and OFDMA). MUD is now evolving from orthogonal (or semi-orthogonal) signaling, toward non-orthogonal signaling, namely, \emph{Non-orthogonal Multiple Access} (NOMA) \referred{NOMAfor5G, NOMAVTC13,UplinkNOMA}\cite{NOMAfor5G, NOMAVTC13,UplinkNOMA}. NOMA aims to better utilize the frequency bands by allowing more users to transmit together in the same frequency at the same time. This paper studies a new NOMA architecture named \emph{Network-Coded Multiple Access} (NCMA).

The key idea of NCMA is to combine physical-layer network coding (PNC) and MUD to enable multipacket reception. PNC, first introduced in \referred{PNC06}\cite{PNC06}, turns mutual interference between signals from simultaneous transmitters to useful network-coded information, thereby improving the throughput of wireless \emph{relay} networks. Most prior PNC works focused on relay networks. NCMA was the first multiple access scheme that explored the use of PNC decoding for non-relay networks, e.g., uplink of wireless local area networks (WLAN) \referred{NCMA1}\cite{NCMA1}.

\begin{figure}
\centering
\includegraphics[width=0.45\textwidth]{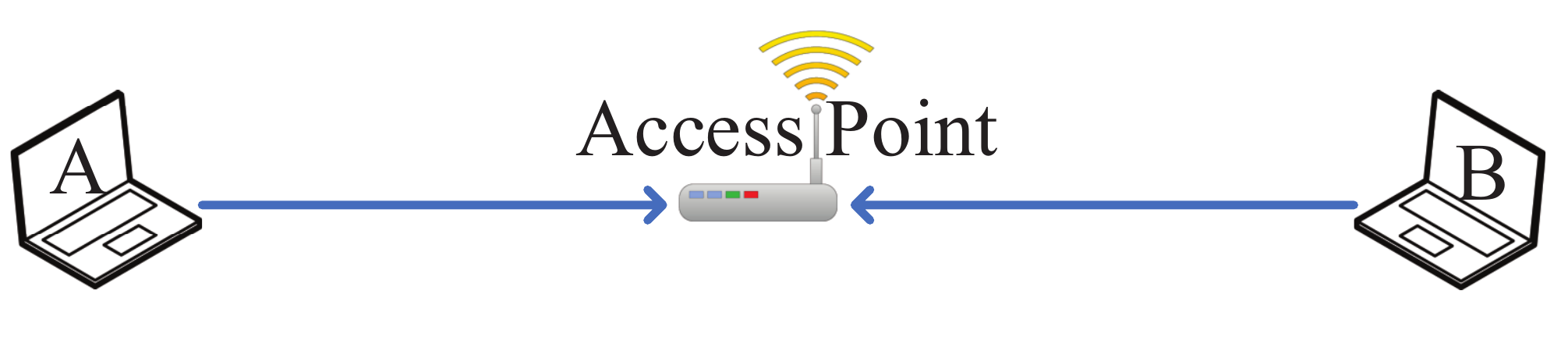}
\caption{System Model for NCMA.}\label{fig:System_model}
\end{figure}

\begin{figure*}[t]
   \begin{minipage}{0.49\linewidth}
     \centering
     \includegraphics[width=0.58\textwidth]{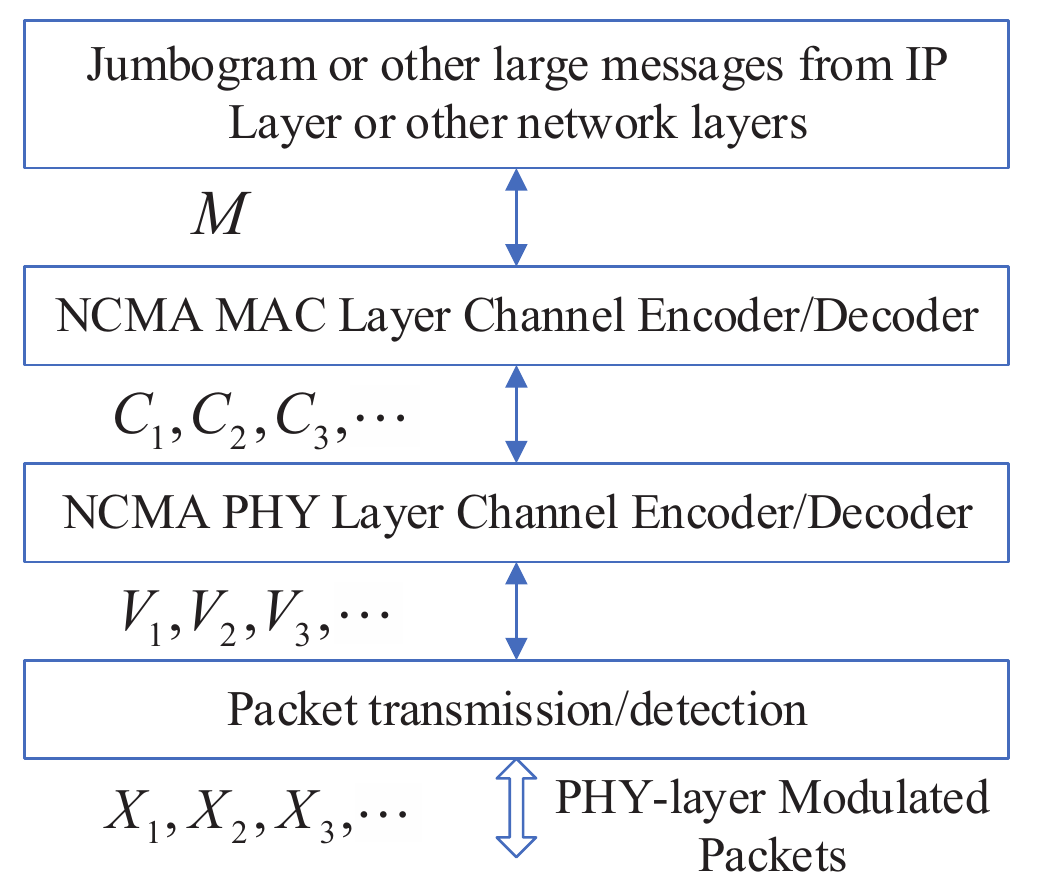}
     \caption{The general architecture of an NCMA node's information processing at MAC and PHY layers.}\label{fig:General_architec}
   \end{minipage}
   \hfill
   \begin{minipage}{0.50\linewidth}
      \centering
      \includegraphics[width=0.88\textwidth]{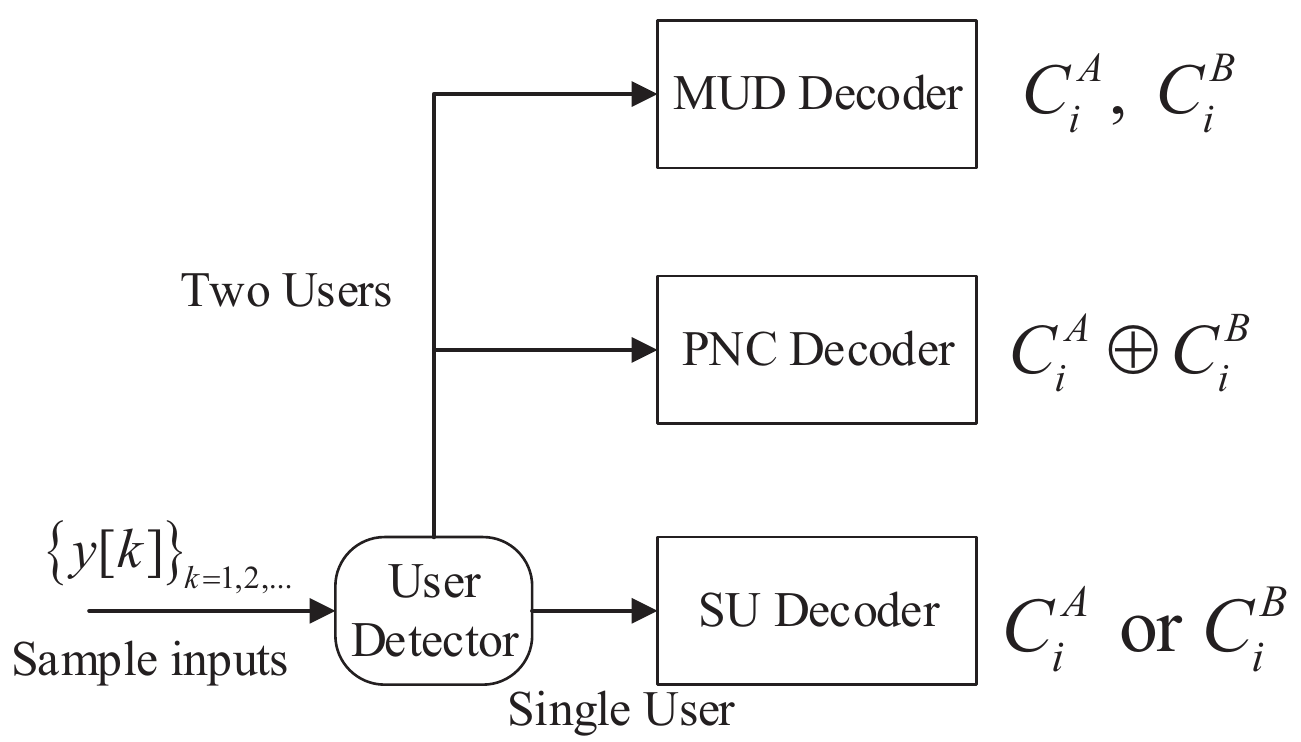}
      \caption{NCMA PHY-layer channel decoders.}\label{fig:NCMA_decoder}
   \end{minipage}
\end{figure*}

Fig. \ref{fig:System_model} shows a typical WLAN setup in which two end nodes send messages to a common access point (AP). The two end nodes are allowed to send their packets simultaneously to boost throughput. NCMA jointly exploits PNC and MUD through a cross-layer design involving channel coding/decoding at the MAC and PHY layers, as explained below. Each client node (e.g., nodes A and B in Fig. \ref{fig:System_model}) partitions and encodes one large source message (e.g., messages $M^A$ and $M^B$ for nodes A and B) into multiple small packets at the MAC layer (see Fig. \ref{fig:General_architec}). At the PHY layer, additional channel coding is performed on each small packet before it is transmitted to the AP. At the AP, two PHY-layer decoders are used to decode useful information from the overlapped packets transmitted simultaneously by different client nodes: (i) the PNC decoder attempts to decode a network-coded packet (e.g., a bit-wise XOR packet $A \oplus B$ \referred{PNC06}\cite{PNC06}), while (ii) the MUD decoder attempts to decode the individual native packets $A$ and $B$. In \referred{NCMA1}\cite{NCMA1}, a two-layer decoding approach was proposed to make use of the PNC packet $A \oplus B$ efficiently. For example, experimental results in \referred{NCMA1}\cite{NCMA1} showed that at SNR of 8.5dB, with probability 22\% the MUD decoder can decode only one of the packet $A$ or $B$. When only one of the two native packets is decoded, with probability 85\% percent the PNC decoder can decode $A \oplus B$. In this scenario, the PHY layer can use $A \oplus B$ and the available native packet to recover the missing native packet, $A$ or $B$. Furthermore, \referred{NCMA1}\cite{NCMA1} showed that when neither packet $A$ nor $B$ is decoded, with probability 40\% the PNC decoder can still decode $A \oplus B$. Such ``lone'' PNC packets are not useful at the PHY layer for the recovery of native packets $A$ and $B$. However, the correlations among successive PHY-layer packets introduced by the MAC-layer channel coding  allows NCMA to make use of the PHY-layer network-coded packets to recover the two MAC-layer native messages $M^A$ and $M^B$. With the two-layer channel coding, NCMA makes good use of the PHY-layer network-coded packets for the recovery of native messages at the MAC layer (see Section \ref{sec:Overview2} for details).

Prior works on NCMA \referred{NCMA1,NCMA2}\cite{NCMA1,NCMA2}, only explored a simple prototype with BPSK modulation. To increase the system throughput, especially in the medium and high signal-to-noise ratio (SNR) regimes, it is desirable to adopt high-order modulations. This paper is an attempt to fill the gap.

It turns out that there are a number of subtleties when applying high-order modulations in NCMA. As will be shown in Section \ref{sec:SingleAntenna}, direct generalization of the scheme in \referred{NCMA1,NCMA2}\cite{NCMA1,NCMA2} from BPSK to QPSK leads to very low system throughput. Rather than a boost, at 8.5dB SNR, the PNC decoder can only successfully decode $A \oplus B$ with probability of 5\%. The system throughput drops by round 80\%, when we move from BPSK to QPSK directly (see Fig. \ref{fig:phy_stat_benchmark} and elaboration in Section \ref{sec:Exp}).

This drastic throughput degradation is caused by the relative phase offset between the signals of simultaneously transmitted packets. Here, we remark that certain advanced channel decoding methods expounded in the literature \referred{liew2015primer, DekorsyWSA2014, WubbenGlobecom10}\cite{liew2015primer, DekorsyWSA2014, WubbenGlobecom10} may partially solve the phase offset problem, but these are generally complex \emph{iterative} decoding methods that induce large decoding latency, and therefore are not amenable to practical implementation.

A goal of ours is to build a high-order modulated NCMA system that can operate in real time. Thus, in this paper, we focus on \emph{non-iterative} Viterbi decoder, assuming the use of convolutional codes. To address the relative phase offset issue, we explore the use of two antennas at the AP $-$ the client nodes still have one antenna each $-$ together with ``\emph{symbol-level}" decoding\footnote{It is worth pointing out that, although our setup shares the same hardware setting with distributed MIMO, a typical MIMO system only incorporates MUD but not PNC decoding at the AP; furthermore, most distributed MIMO systems require transmitter precoding, while our client nodes do not perform transmitter precoding to maintain low complexity of the overall system.}. We refer to our system as \emph{MIMO-NCMA}.


\begin{table*}[t]
\centering
\caption{\textnormal{Related work on iterative and non-iterative decoding schemes for NCMA systems with different modulations (we focus on PNC decoding here, and more MUD schemes are referred to \referred{Verdubook}\cite{Verdubook}).}}
\begin{tabular}{|c|c|c|c|c|}
 \hline
 \backslashbox {\scriptsize{Modulations}}{\scriptsize{Decoding Methods}} &
 {Non-iterative} &
 {Iterative} \\
 \hline
 BPSK & \referred{NCMA1,NCMA2,RPNCSRIF13}\cite{NCMA1,NCMA2,RPNCSRIF13} & \multirow{2}{2.2cm}{\referred{liew2015primer, DekorsyWSA2014, WubbenGlobecom10, BICM-ID-PNC}\cite{liew2015primer, DekorsyWSA2014, WubbenGlobecom10, BICM-ID-PNC}}  \\
 \cline{1-2}
   High-order Modulations  & \referred{longQAM,KoikeJSAC09}\cite{longQAM,KoikeJSAC09} (non-channel-coded)  & \\
   \cline{2-2}
  (QPSK and beyond) &\textbf{Our Work: Joint use of MIMO and symbol-level decoding}  & \\
  \hline
 \end{tabular}
 \label{tab:relatedwork}
\end{table*}

The main results of our investigations can be summarized as follows:

(1) For MIMO-NCMA, we first study a low-complexity demodulation scheme that is compatible with the standard point-to-point Viterbi decoder that admits the soft information of individual bits as inputs. A high-order modulated symbol can be broken into multiple bits as the Viterbi decoder's inputs. We refer to such a decoder as the ``\emph{bit-level}'' NCMA decoder. With two antennas at the AP and the bit-level decoding, the system throughput of QPSK-modulated NCMA can be improved substantially.

(2) Moving from QPSK to 16-QAM (or higher-order modulations), bit-level decoding may lead to severe performance degradation. This is because in NCMA, unlike point-to-point systems, the two users' constellation points (each containing 4 bits under Gray mapping) are correlated within an overlapped symbol, but the bit-level decoders treat the 4 bits as independent information to match the standard point-to-point Viterbi decoder. To avoid information processing loss in the demodulator, this paper puts forth an enhanced PHY-layer decoder for 16-QAM (extendable to higher modulations), referred to as the ``\emph{symbol-level}'' decoder. The symbol-level NCMA decoder contains PNC and MUD demodulators that can retain the information on inter-correlations among the bits inside a symbol. To accommodate the demodulated symbols, rather than bits, a symbol-level Viterbi decoder is applied to accept symbol log-likelihoods. We further show, that our proposed 16-QAM symbol-level Viterbi decoder has exactly the same decoding complexity and the same order of processing time as its bit-level counterpart. 

For performance evaluation, we implemented the bit-level and symbol-level PHY-layer decoders on software-defined radio. Our experiments show that, at 10dB SNR, the throughput of QPSK MIMO-NCMA is higher than those of conventional MIMO MUD systems operated with distributed ZF and MMSE decoders by 100\% and 80\%, respectively. More importantly, our experimental results show that, at SNR=10dB, the throughput of our QPSK MIMO-NCMA is double that of both the prior BPSK NCMA. At SNR=20dB when 16-QAM can be supported, the throughput of MIMO-NCMA can be as high as 3.5 times that of the prior BPSK NCMA. Overall, we provide an implementable framework for high-order modulated NCMA.

The remainder of this paper is organized as follows: Section \ref{sec:RelatedWork} overviews prior related work. Section \ref{sec:Overview} describes the key idea of the NCMA system. Section \ref{sec:SingleAntenna} studies the penalty induced by relative phase offsets in high-order modulations. Following that, Section \ref{sec:MIMO_NCMA} puts forth our solutions. Section \ref{sec:Symbol-level_NCMA} introduces enhanced PHY-layer decoders. Section \ref{sec:Exp} presents the implementation details of our approach and the associated experimental results. Finally, Section \ref{sec:Conclusions} concludes this paper.

\section{Related Work} \label{sec:RelatedWork}
\emph{\textbf{Physical-layer Network Coding}}: Ref. \referred{PNC06}\cite{PNC06} first proposed PNC to increase the throughput of a two-way relay network (TWRN). In TWRN, two end nodes exchange information via a relay. PNC doubles the throughput of a TWRN operated with the traditional scheduling scheme \referred{PNC06}\cite{PNC06}. PNC has been studied and evaluated in depth: we refer the interested readers to \referred{liew2015primer,popovski2006anti,Nazer2011ReliablePNC,FPNCPhycom12,RPNCSRIF13}\cite{liew2015primer,popovski2006anti,Nazer2011ReliablePNC,FPNCPhycom12,RPNCSRIF13} and the references therein for details. Following the tradition of \referred{PNC06}\cite{PNC06}, prior PNC work focused almost exclusively on \emph{relay} networks. By contrast, NCMA was the first attempt to apply PNC in \emph{non-relay} networks (i.e., multiple access in wireless networks) \referred{NCMA1,NCMA2}\cite{NCMA1,NCMA2}. There has been existing work focusing on high-order modulated PNC \referred{BICM-ID-PNC,wang2012constellation,yang2010modified}\cite{BICM-ID-PNC,wang2012constellation,yang2010modified}. This work, however, was theoretical in nature and assumed perfect synchronization (e.g., no phase offset between the concurrent transmitting signals). The practical implementation of such  synchronized systems is much more challenging than the current system studied in our paper.

The phase offset issue in high-order modulated PNC has also been studied and partially solved through different decoding schemes, as shown in Table \ref{tab:relatedwork}. Unlike the existing work in Table \ref{tab:relatedwork}, the solution of using multiple antennas and symbol-level decoding in our paper provides a non-iterative decoding scheme that is simple to implement.

\emph{\textbf{Coding for Multiple Access Channels}}: Besides NCMA \referred{NCMA1,NCMA2}\cite{NCMA1,NCMA2}, there have been other efforts to apply network coding in multiple access networks. The major difference between NCMA and this work is that NCMA tries to decode more than one equation per reception (e.g., tries to decode both the PNC packet and the individual native packets) from one overlapped packet, while most other work targets to get one equation per reception (either the PNC packet or one native packet). For example, \referred{RAwithPNC,Cocco2011,SeekandDecode}\cite{RAwithPNC,Cocco2011,SeekandDecode}  explored forming linear equations from the collided packets to derive source packets. Refs. \referred{LivaGraph,CodedRandomAccess}\cite{LivaGraph,CodedRandomAccess} treated the collisions in multiple access channel as erasure-correcting codes on graphs. But \referred{Cocco2011,LivaGraph,CodedRandomAccess}\cite{Cocco2011,LivaGraph,CodedRandomAccess} only form one equation for each overlapped packet, whereas NCMA can form one or two equations for each overlapped packet depending on the instantaneous channel condition. Furthermore, the decoding in \referred{RAwithPNC, SeekandDecode}\cite{RAwithPNC, SeekandDecode} is based on PHY-layer equations only, while NCMA makes use of another MAC-layer channel coding based on the PHY-layer PNC packets.

Decoding source packets from concurrent transmissions, e.g., Non-orthogonal Multiple Access (NOMA), were also studied in \referred{NOMAfor5G, Strider2011,NOMAVTC13,UplinkNOMA,AutoMAC12}\cite{NOMAfor5G, Strider2011,NOMAVTC13,UplinkNOMA,AutoMAC12}, where successive interference cancellation (SIC) or other MUD schemes were adopted for multipacket reception. Again, none of them considered PNC decoding. A general comparison between NOMA and traditional TDMA, FDMA, CDMA schemes can be found in \referred{NOMAfor5G}\cite{NOMAfor5G}.

\begin{figure*}[t]
   \begin{minipage}{0.45\linewidth}
     \centering
     \includegraphics[width=0.65\textwidth]{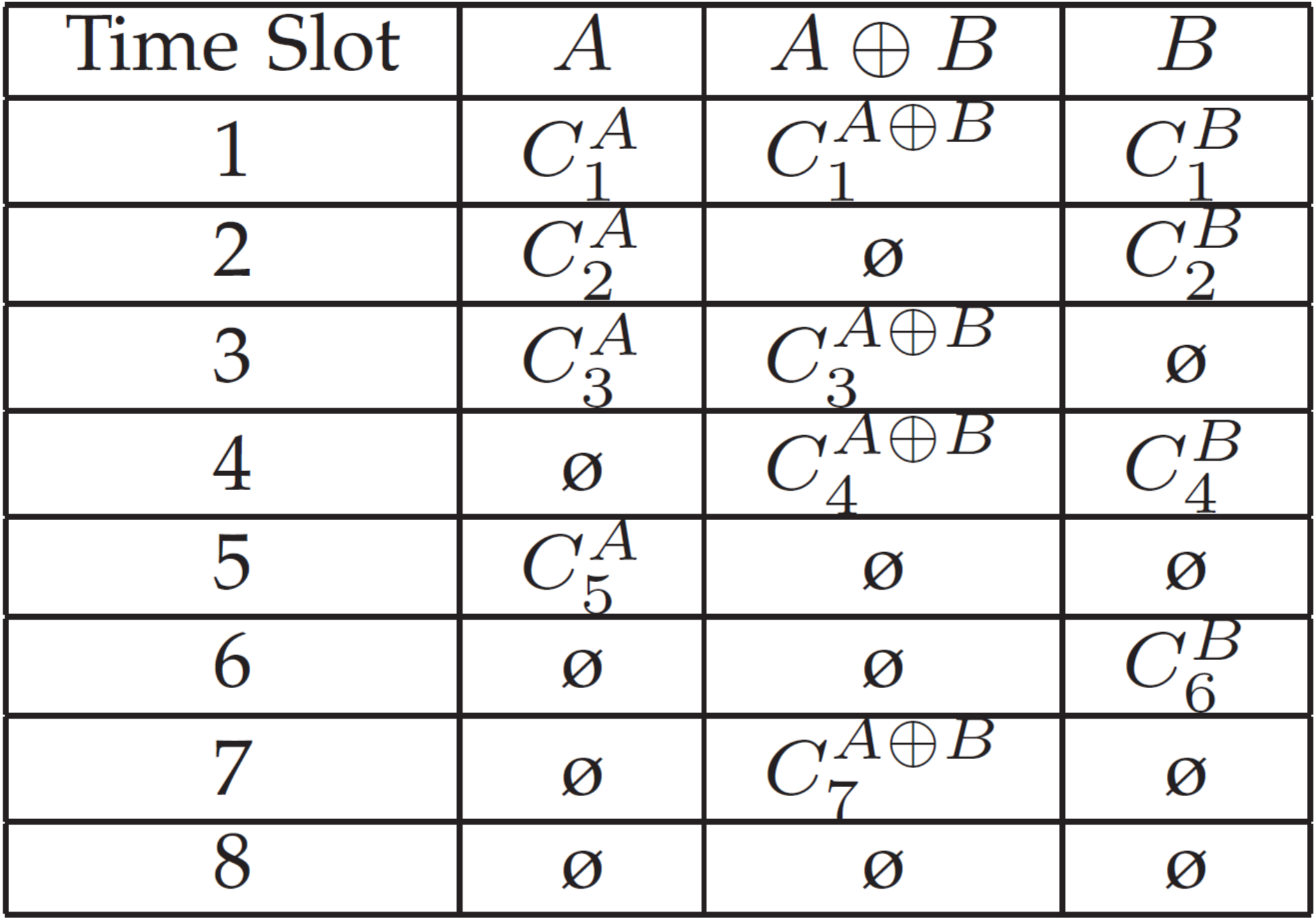}
     \caption{An example of PHY-layer packet reception patterns for concurrently transmitted packets using PNC and MUD decoders. $\emptyset$ means the corresponding packets cannot be decoded.}\label{fig:PHY-bridging}
   \end{minipage}
   \hfill
   \begin{minipage}{0.51\linewidth}
      \centering
      \includegraphics[width=1.1\textwidth]{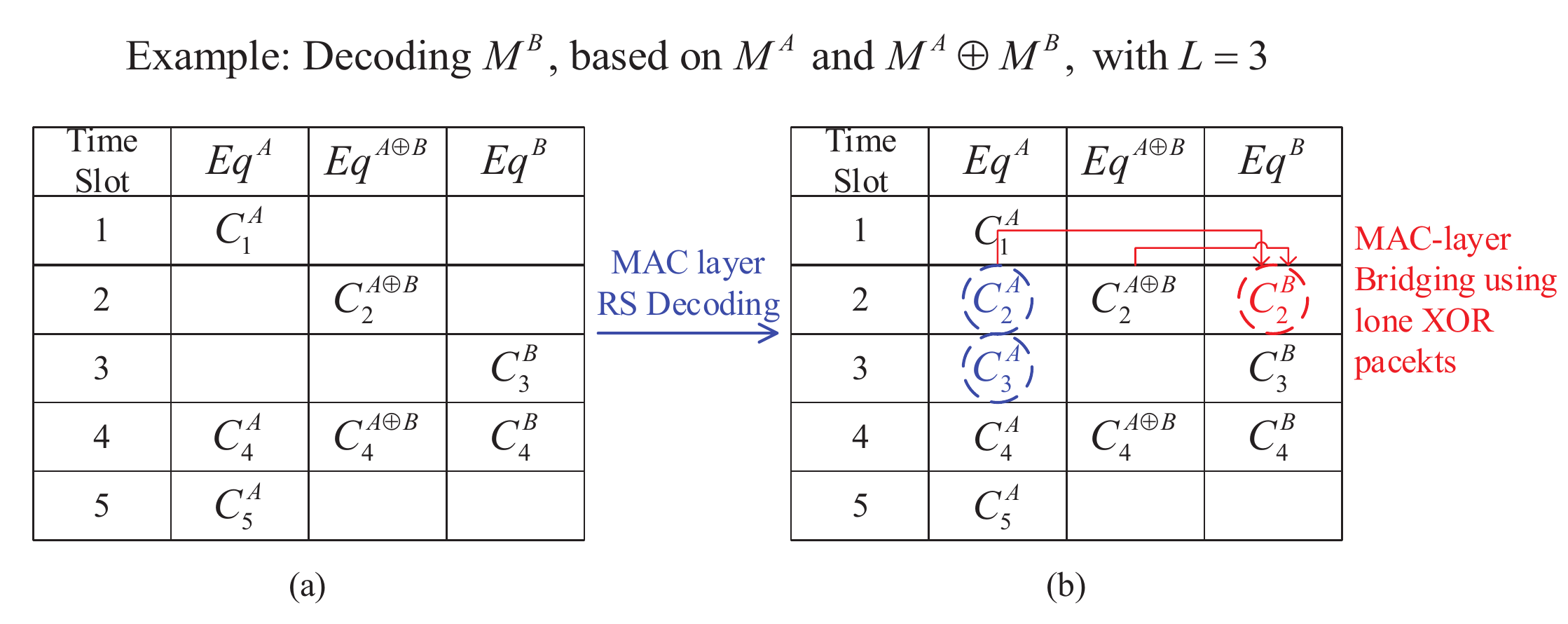}
      \caption{NCMA MAC-layer decoding and bridging example, using L=3 RS code: (a) The decoding outcomes after PHY-layer bridging; (b) MAC-layer RS decoding and bridging using lone XOR packets.}\label{fig:MAC-bridging}
   \end{minipage}
\end{figure*}

\emph{\textbf{Distributed MIMO}}: As with distributed MIMO systems, in MIMO-NCMA, the AP also has multiple antennas. Distributed MIMO can enable spatially separated transmitters to form a virtual MIMO system for multiple access. In the literature, \referred{SAM2009, JMB2012, BalanAirSyncToN2013}\cite{SAM2009,JMB2012,BalanAirSyncToN2013} studied distributed MIMO systems to increase the system throughput, and we refer to them as MIMO-MUD since they only focus on MUD decoding, without incorporating PNC decoding. In this paper, we consider classical MIMO-MUD with zero-forcing (ZF), minimum mean square error (MMSE) decoders as our benchmarks. More sophisticated distributed MIMO decoders can be found in \referred{tse2005fundamentals}\cite{tse2005fundamentals}.

\emph{\textbf{Symbol-level Decoding in Point-to-point Systems}}: The idea of symbol-level decoding can also be used in conventional point-to-point systems. However, unlike in NCMA, the motivation is lacking there because bit-level decoding already yields performance that approaches the limit of the Shannon capacity. For example, the Gray-mapped bit-interleaved coded modulation (BICM) point-to-point systems (i.e., a bit-level decoding) have previously been shown to be capacity optimal even without iterative decoding, if the bit positions in the symbol are independent \referred{BICM,GrayBICM}\cite{BICM,GrayBICM}. In NCMA, however, the symbol-level decoding can lead to large decoding improvement over the bit-level counterpart because of the correlations among bits within the overlapped symbol. The detailed comparisons between NCMA symbol-level and bit-level decodings will be presented in Section \ref{sec:MIMO_NCMA3} and \ref{sec:Exp22}, including the simulation and experimental results, respectively.

\section{NCMA Overview} \label{sec:Overview}
\subsection{General System Model for NCMA} \label{sec:Overview1}
We study a multiple access system where two end nodes, A and B, transmit information to an access point (AP) simultaneously, as shown in Fig. \ref{fig:System_model}. We consider the use of both physical-layer network coding (PNC) and multiuser decoding (MUD) to boost system throughput. This system is referred to as a network-coded multiple access (NCMA) system \referred{NCMA1}\cite{NCMA1}.

NCMA includes both MAC layer and PHY layer operations. With respect to Fig. \ref{fig:General_architec}, at the MAC layer, a large message $M^A$ of node A is divided and encoded into multiple packets, $C_i^A, i=1,2,...$ Similarly, a large message $M^B$ of node B is encoded into multiple packets, $C_i^B, i=1,2,...$ We assume the use of the Reed-Solomon (RS) code when coding a large message into multiple packets. At the PHY layer, each packet  $C_i^A$ (or $C_i^B$) is further channel-encoded into $V_i^A$ ($V_i^B$) for reliable transmission. We adopt the convolutional code as the PHY-layer channel codes. NCMA is a time-slotted system. That is, each end node $j$ transmits packets  $V_1^j,V_2^j,...,V_i^j,...$  to the AP in successive time slots, and the two end nodes' packets (i.e.,  $V_i^A$ and $V_i^B$) are configured to transmit simultaneously in the same time slot $i$.

In the uplink transmission of NCMA, at the PHY layer, as shown in Fig. \ref{fig:NCMA_decoder}, the AP first detects how many nodes are transmitting. When only one node is transmitting, the Single-User (SU) decoder will be used. When two nodes are transmitting simultaneously, the AP receives signals containing two superimposed packets. In this case, the AP decodes using two decoders: the MUD decoder and the PNC decoder. The MUD decoder attempts to decode both packets  $C_i^A$ and $C_i^B$ explicitly, and the PNC decoder attempts to decode a linear combination\footnote{In this paper, we only consider the bit-wise eXclusive-OR (XOR), $ \oplus $, operation of $C_i^A$ and $C_i^B$.} of two packets $C_i^A$ and $C_i^B$, i.e., $C_i^A \oplus C_i^B$. The successfully decoded packets from the PHY layer in different time slots are collected and passed to the MAC layer for further processing. With the help of the MAC-layer RS code, the AP decodes the original messages ${M^A}$ and  ${M^B}$, as elaborated in Part B below. .

\subsection{An Example} \label{sec:Overview2}
Different from traditional multipacket reception systems where only MUD was adopted \referred{Verdubook}\cite{Verdubook}, a main distinguishing feature of NCMA is that it combines PNC decoding with MUD to improve the system throughput. In particular, it is possible that sometimes only $C_i^A \oplus C_i^B$ can be decoded using PNC decoding while MUD fails to recover either $C_i^A$ or $C_i^B$. In this subsection, we illustrate the advantages and the key idea of NCMA using a simple example.

\emph {\textbf{PHY-layer Bridging}} $-$ Let us focus on the PHY-layer decoding outcomes first. In a time slot $i$, for the MUD decoder, there are four possible outcomes: (i) both $C_i^A$ and $C_i^B$  are successfully decoded; (ii) only $C_i^A$ is successfully decoded; (iii) only $C_i^B$ is successfully decoded; (iv) neither $C_i^A$ nor $C_i^B$ can be decoded. For the PNC decoder, there are two possible outcomes: (a)  $C_i^A \oplus C_i^B$ is successfully decoded; (b) $C_i^A \oplus C_i^B$ cannot be decoded. As a result, we have  $4 \times 2 = 8$ possible outcomes. For explanation purposes, Fig. \ref{fig:PHY-bridging} shows an example in which the eight possible outcomes occur in eight successive time slots. In time slots 3 and 4, $C_3^A$ and $C_3^A \oplus C_3^B$ (abbreviated as $C_3^{A \oplus B}$),  $C_4^B$ and  $C_4^A \oplus C_4^B$ (abbreviated as $C_4^{A \oplus B}$) are decoded, respectively. The ``complementary'' XOR packets $C_3^{A \oplus B}$ and $C_4^{A \oplus B}$  can be used to recover individual missing packets $C_3^B$  and $C_4^A$. This process, which leverages the complementary XOR packets, is referred to as PHY-layer bridging \referred{NCMA1}\cite{NCMA1}.

\emph {\textbf{MAC-layer Bridging}} $-$ In Fig. \ref{fig:PHY-bridging}, PHY-layer bridging cannot be applied to time slot 7 because neither native packet  $C_7^A$ nor $C_7^B$  is available, and only a ``lone'' network-coded packet  $C_7^{A \oplus B}$ (namely, PNC packet) is decoded. In NCMA, such lone PNC packets turn out to be useful in MAC-layer decoding. Fig. \ref{fig:MAC-bridging} gives an example illustrating the main idea. Fig. \ref{fig:MAC-bridging}(a) shows the PHY-layer decoding outcomes for a number of successive time slots. We assume the AP has recovered enough native packets  $C_i^A$ to decode $M^A$  with the help of the MAC-layer RS code by time slot 5 $-$ in this example, $L=3$ PHY-layer packets are needed to recover $M^A$. With MAC-layer decoding, native packets $C_2^A$ and $C_3^A$ can also be decoded (conceptually, we could obtain  $C_2^A$ and $C_3^A$  based on re-encoding the recovered  $M^A$ at the MAC-layer, although in practice, a simpler process is possible). Note that the PHY layer failed to obtain $C_2^A$ in time slot 2, but the MAC layer recovers it in time slot 5. With $C_2^A$, the original lone PNC packet $C_2^A \oplus C_2^B$ in time slot 2 becomes a complementary XOR packet. Consequently, we can recover $C_2^B$ (using $C_2^A$ and $C_2^A \oplus C_2^B$), and therefore node B now has enough native packets (i.e., $L=3$) to recover message, as shown in Fig. \ref{fig:MAC-bridging}(b). We refer this process as MAC-layer bridging\referred{NCMA1,NCMA2}\cite{NCMA1,NCMA2}.

\section{Single Antenna System}\label{sec:SingleAntenna}

Despite the throughput improvement brought about by PHY-layer bridging and MAC-layer bridging, the previous NCMA systems\referred{NCMA1,NCMA2}\cite{NCMA1,NCMA2} operate with BPSK modulation only. This upper bounds the throughput to two decoded bits per channel use (one decoded bit from one user). At medium to high signal-to-noise ratio (SNR), e.g., SNR$\geq$10dB, it is desirable to use high-order modulations to further boost the throughput. This paper considers high-order modulations to avoid the saturation of data rate \referred{HalperinSNR10}\cite{HalperinSNR10}.

\begin{figure}[t]
    \centering
    \includegraphics[width=0.5\textwidth]{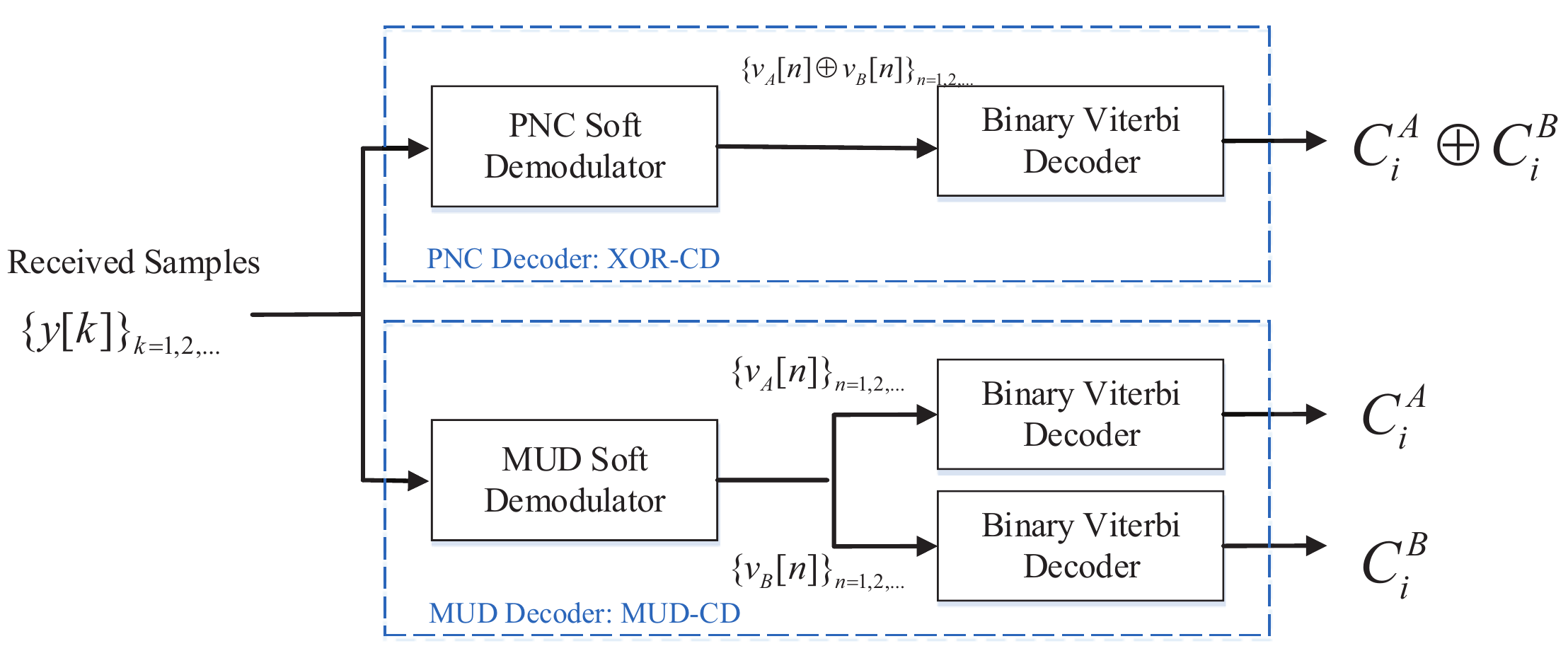}
    \caption{NCMA PHY-layer PNC decoder (XOR-CD) and MUD decoder (MUD-CD) for simultaneous transmissions from two nodes.}\label{fig:PHY-decoder}
\end{figure}

In \referred{NCMA2}\cite{NCMA2}, both PHY and MAC layer real-time decoders have been evaluated on the software-defined radio platform. At the PHY layer, the PNC decoder is an XOR-CD decoder and the MUD decoder is an MUD-CD decoder. XOR-CD first demodulates the overlapped signals into the XOR of the channel-coded bit pairs of A and B. After that, Channel Decoding is applied to obtain the XOR of the source bit-pairs of A and B.  MUD-CD, on the other hand, first  demodulate the overlapped signals into separate channel-coded bits of A and B. After that, Channel Decoding is applied on each of the streams to obtain the source bits of A and B (details of XOR-CD and MUD-CD can be found in the Parts A and B below). A salient feature of these decoders is that the standard low-complexity point-to-point binary Viterbi channel decoder can be used with changes on the demodulators only, as shown in Fig. \ref{fig:PHY-decoder}. However, as will be elaborated, both PNC and MUD decoders encounter a critical ``phase offset'' problem when we move from BPSK to high-order modulations. In the following, we illustrate this problem in XOR-CD and MUD-CD assuming QPSK.

\subsection{Phase Penalty for PNC Decoder} \label{sec:SingleAntenna1}

Several decoding approaches are possible for channel-coded PNC systems \referred{liew2015primer}\cite{liew2015primer}. In this paper, since we aim for real-time operations rather than optimal performance, we use the simple XOR-CD decoder as the PNC decoder. Sophisticated PNC decoders with better BER performance are possible, at the cost of high computational complexity and large decoding delays, e.g., Jt-CNC (joint channel-decoding and network-coding) \referred{liew2015primer}\cite{liew2015primer}. They have been studied in the literature, and we refer interested readers to \referred{liew2015primer, WubbenGlobecom10}\cite{liew2015primer, WubbenGlobecom10} for further details.

The general architecture of XOR-CD is shown in the upper block of Fig. \ref{fig:PHY-decoder}. Let ${V^A} = ({v_A}[1],...,{v_A}[n],...)$ denote the PHY-layer codeword of node A in one time slot (i.e., one binary encoded packet), where ${v_A}[n]\in\{ 0,1\}$ is the $n$-th convolutional encoded bit (similarly, we have ${V^B} = ({v_B}[1],...,{v_B}[n],...)$ for node B). Assuming QPSK modulation, the $k$-th modulated symbol ${x_A}[k]$ for the PHY-layer transmitted packet ${X^A} = ({x_A}[1],...,{x_A}[k],...)$ is given by
\begin{align}
{x_A}[k] = (1 - 2{v_A}[2k - 1]) + (1 - 2{v_A}[2k])j, ~k = 1,2,3,...
\label{equ:x_A}
\end{align}
Let us assume an OFDM system where multipath fading can be dealt with by the cyclic prefix (CP). The $k$-th received sample in the frequency domain at the AP can be written as
\begin{align}
{y_R}[k] = {h_A}[k]{x_A}[k] + {h_B}[k]{x_B}[k] + w[k],
\label{equ:y_R}
\end{align}

\noindent where $w[k]$ is the noise term, and ${h_A}[k]$, ${h_B}[k]$ are the channel gains of the $k$-th samples of nodes A and B, respectively. In XOR-CD, the received samples ${\left\{ {{y_R}[k]} \right\}_{k = 1,2,3,...}}$ are first passed through the PNC demodulator to obtain the XOR bits ${\{ v_A[n] \oplus v_B[n]\} _{n = 1,2,...}}$. Note that the outputs ${\{ v_A[n] \oplus v_B[n]\} _{n = 1,2,...}}$ from the PNC demodulator can be hard or soft bits (i.e., log likelihood ratio of the bits). The decoders implemented by us in this work, and the BER performance results therein, assume the use of soft bits. These bits are then fed to a standard binary Viterbi decoder (as used in a point-to-point system) to decode the network-coded packet $C_i^A \oplus C_i^B$. The standard Viterbi decoder can be used for PNC decoding because XOR-CD exploits the linearity of linear channel codes, such as convolutional codes. Specifically, define $\Pi ( \cdot )$ as the convolutional channel encoder. Since $\Pi ( \cdot )$ is linear, we have $ V^A \oplus V^B = \Pi \left( {C^A} \right) \oplus \Pi \left( {C^B} \right) = \Pi \left( {C^A \oplus C^B} \right)$.

With respect to node A, the odd (even) bits of $V^A$ are mapped to the in-phase (quadrature) part of ${x_A}[k]$ in QPSK, i.e., $x_A^I[k] = 1 - 2{v_A}[2k - 1]$ ($x_A^Q[k] = 1 - 2{v_A}[2k]$). A particular pair of symbols from the two nodes is expressed as $({x_A}[k],~{x_B}[k]) = (x_A^I[k] + x_A^Q[k]j,~x_B^I[k] + x_B^Q[k]j)$.

An important issue in PNC is how to calculate ${x_A}[k] \oplus {x_B}{\rm{[k] }}$ (abbreviated as ${x_{A \oplus B}}[k]$) using the received sample ${y_R}[k]$ in (\ref{equ:y_R}). To maintain the linear property of convolutional codes, in XOR-CD we need to map the in-phase part of ${x_A}[k]$ with the in-phase part of ${x_B}[k]$, i.e., $x_A^I[k]$ with $x_B^I[k]$  (similarly, $x_A^Q[k]$ with $x_B^Q[k]$) into the network-coded in-phase (quadrature) part of ${x_{A \oplus B}}[k]$. More precisely, the PNC mapping in XOR-CD for ${x_{A \oplus B}}[k]$ is defined as
\begin{align}
{x_{A \oplus B}}[k] = x_A^I[k] \oplus x_B^I[k] + (x_A^Q[k] \oplus x_B^Q[k])j,
\label{equ:xor-mapping}
\end{align}

\noindent where $x_A^I[k]\oplus x_B^I[k] = x_A^I[k]x_B^I[k]$ and $x_A^Q[k]\oplus x_B^Q[k] = x_A^Q[k]x_B^Q[k]$ given that $x_A^I[k]$, $x_B^I[k]$, $x_A^Q[k]$, $x_B^Q[k] \in \{ 1, - 1\}$. The PNC demodulation rule for the XORed bits is defined as
\begin{align}
{v_A}[2k - 1] \oplus {v_B}[2k - 1] = \frac{{1 - x_A^I[k] \oplus x_B^I[k]}}{2}, \notag \\
{v_A}[2k] \oplus {v_B}[2k] = \frac{{1 - x_A^Q[k] \oplus x_B^Q[k]}}{2}.
\label{equ:pnc_demo}
\end{align}

\begin{figure}[!t]
    \centering
    \includegraphics[width=0.42\textwidth]{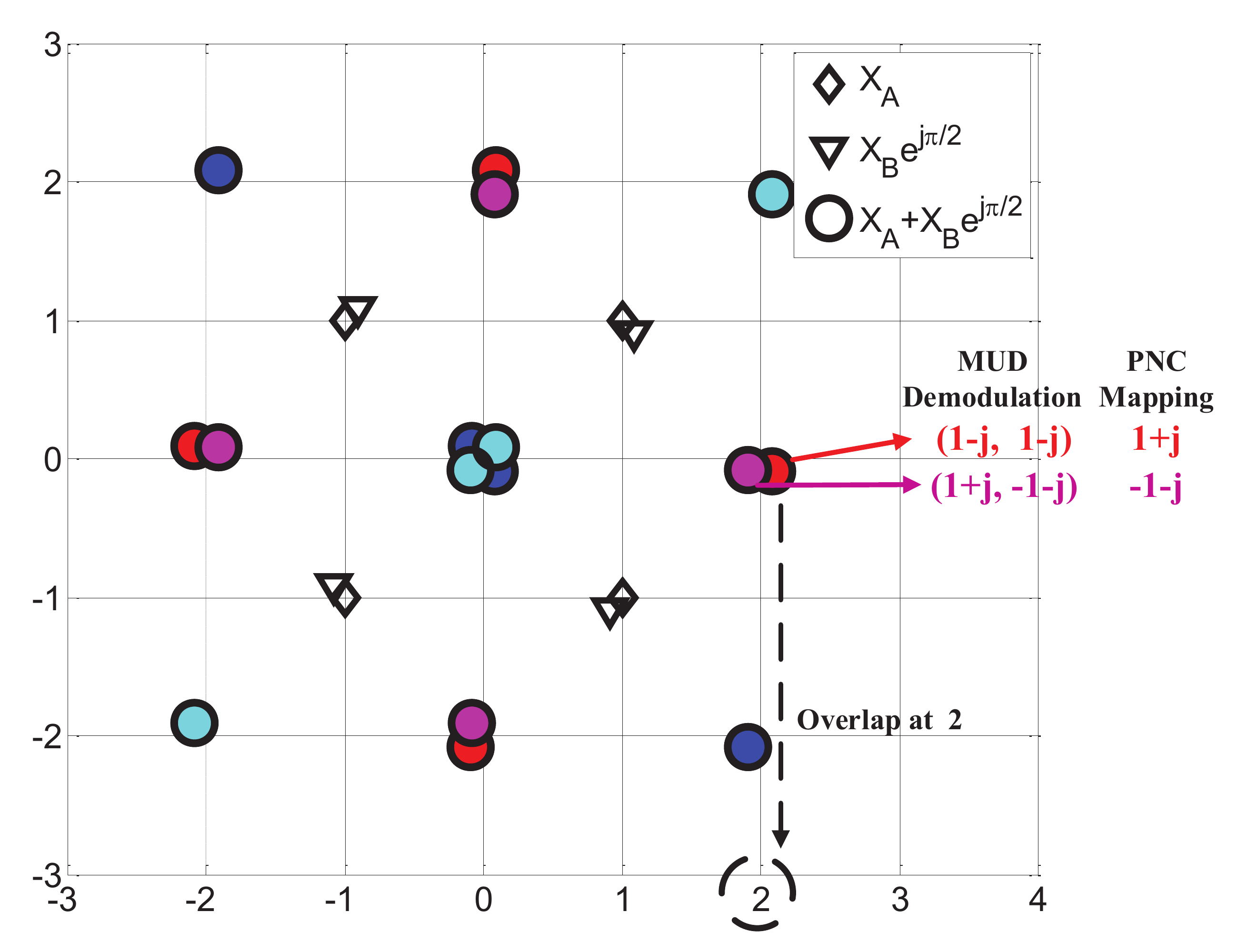}
    \caption{Constellation map for the received samples at the AP for $\left| {{h_A}} \right| = \left| {{h_B}} \right| = 1$ and relative phase offset $\Delta \phi = \pi/{2}$. Note that we purposely set $\Delta \phi$ to be slightly smaller than $\pi/{2}$ to highlight different PNC mappings using different colors, where the same color corresponds to the same network-coded symbol. The symbol pair $({x_A},{x_B})$ denotes the MUD demodulated symbols.}\label{fig:Constellation}
\end{figure}

\begin{figure*}[t]
     \centering
     \includegraphics[width=0.99\textwidth]{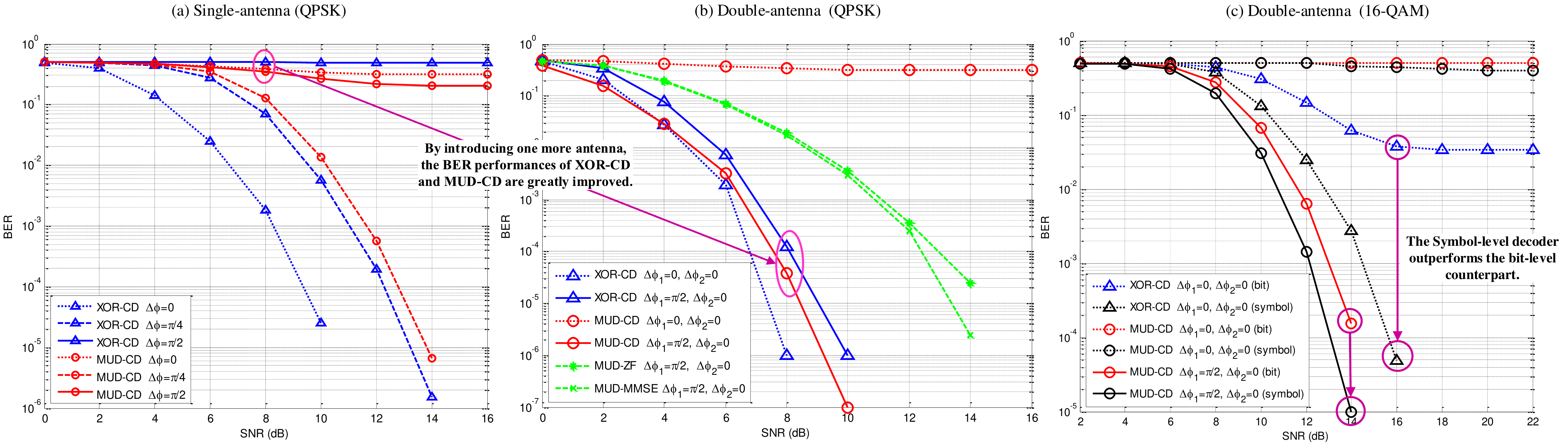}
     \caption{BER results for single-antenna and double-antenna NCMA systems with different relative phase offsets in AWGN channels: XOR-CD and MUD-CD decoders with (a) single antenna, (b) double antennas with QPSK, benchmarked by ZF and MMSE decoders, and (c) double antennas with 16-QAM.}
\label{fig:Ber_curve}
\end{figure*}

Let us focus on one particular received sample ${y_R}[k] = {x_A}[k] + {x_B}[k]{e^{j\Delta \phi }}$, where $\Delta \phi$ is the relative phase offset between the two nodes (to explain things in simple terms, here we assume perfect power control so that the received signal powers for both nodes are equal, and we have ${h_A}[k] = 1$ and ${h_B}[k] = {e^{j\Delta \phi }}$; in general, this needs not be the case). Fig. \ref{fig:Constellation} plots the noise-free constellation map for $\Delta \phi  = \pi / {2}$, in which some constellation points overlap with the others (note: to see the overlapping constellations points more clearly in the figure, we purposely set $\Delta \phi$ to be slightly smaller than $\pi / {2}$).  In Fig. \ref{fig:Constellation}, constellation points of the same color are mapped to the same XOR value. Note, for example, that the constellation points of symbol pairs $(1-j,1-j)$ and $(1+j,-1-j)$  overlap, but they are mapped to different XOR values. In particular, when $\Delta \phi  = \pi / {2}$, the XOR mapping of (\ref{equ:xor-mapping}) leads to ambiguity even in the absence of noise, and the error probability for the network-coded symbol can be as high as 50\%.

From the simulation results in Fig. \ref{fig:Ber_curve}(a), we can see that the BER performance of XOR-CD with $\Delta \phi  = \pi / {2}$ degrades greatly compared with that of $\Delta \phi  = \pi / {4}$. In general, we find that the BER performance of XOR-CD depends much on the relative phase offset $\Delta \phi$. As will be discussed in the next subsection, the BER performance of MUD-CD is also highly correlated with the relative phase offset $\Delta \phi$.

\subsection{Phase Penalty for MUD Decoder} \label{sec:SingleAntenna2}
The architecture of MUD-CD is shown in the lower block of Fig. \ref{fig:PHY-decoder}. The goal of the MUD-CD decoder is to decode the two source packets $C_i^A$ and $C_i^B$ separately. In this process, the received samples ${\left\{ {{y_R}[k]} \right\}_{k = 1,2,3,...}}$ are first passed through a MUD demodulator to get the binary channel-encoded bits ${\{ v_A[n]\} _{n = 1,2,...}}$ and ${\{ v_B[n]\} _{n = 1,2,...}}$, which are then fed into two binary Viterbi decoders to recover the packets $C_i^A$ and $C_i^B$ of nodes A and B, respectively.

Phase penalty similar to that in XOR-CD also exists in MUD-CD, as shown in Fig. \ref{fig:Ber_curve}(a). Let us use the constellation map in Fig. \ref{fig:Constellation} to explain the phase penalty problem for MUD-CD. With respect to a particular constellation point ``2'', we cannot distinguish between the symbol pair $(1-j,1-j)$ and symbol pair $(1+j,-1-j)$, based on the received sample ${y_R}[k]$ when $\Delta \phi  = \pi / {2}$. We find that both PNC and MUD decoders' BER performances degrade drastically when $\Delta \phi  = \pi / {2}$ (see  Fig. \ref{fig:Ber_curve}(a)).

From the BER curves in Fig. \ref{fig:Ber_curve}(a), we can see that the BER performance of XOR-CD is also related to the relative phase offset $\Delta \phi$ between two nodes. We find that both PNC and MUD decoders' BER performances degrade drastically when $\Delta \phi  = \pi / {2}$. In general, when $\Delta \phi$ is in the range of $\pi / {4} \le \Delta \phi  \le 3\pi / {4}$ or $5\pi / {4} \le \Delta \phi  \le 7\pi / {4}$, XOR-CD has poor performance; when $\Delta \phi  = {m\pi}/2,m =0,1,2,...$, the overlapping constellation points degrade the performance of MUD-CD decoder. This is a hurdle for NCMA systems when QPSK is adopted. This hurdle can be overcome with the use of multiple antennas at the AP, as explained in Section \ref{sec:SingleAntenna3}.


\subsection{Possible Solution to Alleviate Phase Penalty}\label{sec:SingleAntenna3}

The fundamental reason why the BER performance of NCMA is bad when $\Delta \phi  = \pi / {2}$ is that some overlapping constellation points are being mapped to different network-coded symbols in the case of PNC decoding, and can be demodulated into two different symbol pairs in the case of MUD (i.e., there is an ambiguity even in the absence of noise). In the literature, several approaches have been proposed to partially solve the phase penalty problem for PNC systems. For example, \referred{liew2015primer}\cite{liew2015primer} proposed a real-to-imaginary PNC mapping for ${x_{A \oplus B}}$ (i.e., $x_A^I \oplus x_B^Q + (x_A^Q \oplus x_B^I)j$ rather than using (\ref{equ:xor-mapping})) when the phase offset is $\pi / {4} \le \Delta \phi  \le 3\pi / {4}$ when the data are not channel-coded.  However, this method is not applicable to channel-coded PNC systems because the linearity of channel codes will be destroyed by such PNC mapping (i.e., the PNC-mapped symbols do not constitute a valid codeword anymore). Another possible method was put forth by \referred{KoikeJSAC09}\cite{KoikeJSAC09}, in which a QPSK to 5-QAM demodulation was adopted. This method, however, is also for non-channel-coded systems and it cannot be easily extended to channel-coded XOR-CD systems. Refs. \referred{liew2015primer}\cite{liew2015primer} studied a high computational-complexity Jt-CNC (joint channel-decoding and network-coding) approach that makes real-time processing difficult.

Nowadays, many APs are equipped with multiple antennas \referred{dot11std13}\cite{dot11std13}. In this paper, we consider an NCMA system in which the AP has two antennas, referred to as \emph{MIMO-NCMA}. We show via simulations and experiments that MIMO-NCMA can solve the phase penalty problem while maintaining the low complexity of non-iterative PHY-layer decoders. 

In Section \ref{sec:MIMO_NCMA}, we show that MIMO techniques can solve the phase penalty issue in QPSK-modulated NCMA systems, using bit-level decoding as in point-to-point systems.

In Section \ref{sec:Symbol-level_NCMA}, we show that just using the MIMO and bit-level decoding techniques may not be good enough when higher-order modulations beyond QPSK are adopted, e.g., 16-QAM. Fortunately, replacing bit-level decoding with symbol-level decoding greatly improves the performance. 

\section{MIMO-NCMA: QPSK with Bit-level Decoding}\label{sec:MIMO_NCMA}
This section presents bit-level PHY-layer decoders for MIMO-NCMA with QPSK modulations: Section \ref{sec:MIMO_NCMA1} focuses on the design of the XOR-CD decoder, and Section \ref{sec:MIMO_NCMA2}, the MUD-CD decoder. For the PHY-layer channel codes, our system adopts the same $[133, 171]_8$ convolutional code as in the 802.11 standard \referred{dot11std13}\cite{dot11std13}. We present a low-complexity demodulation scheme that is compatible with the standard point-to-point bit-level Viterbi decoder that admits individual bits' information as inputs, e.g., a QPSK symbol is broken into two bits as the Viterbi decoder's inputs. 

\subsection{PNC Decoder (XOR-CD)} \label{sec:MIMO_NCMA1}
Let the received samples on the two antennas at the AP be ${\left\{ {{y_{R1}}[k]} \right\}_{k = 1,2,3,...}}$ and ${\left\{ {{y_{R2}}[k]} \right\}_{k = 1,2,3,...}}$, respectively. Our target is to compute the log-likelihood ratios (LLR) of two bits ${v_A}[2k - 1] \oplus {v_B}[2k - 1]$ and ${v_A}[2k] \oplus {v_B}[2k]$ based on the $k$-th received samples ${y_{R1}}[k]$ and ${y_{R2}}[k]$ of antennas 1 and 2. The PNC demodulator's outputs (namely, the soft information of ${\{ v_A^{}[n] \oplus v_B^{}[n]\} _{n = 1,2,...}}$) are fed into the Viterbi decoder. This subsection derives the soft information of ${\{ v_A^{}[n] \oplus v_B^{}[n]\} _{n = 1,2,...}}$.
We assume the end nodes use QPSK modulation and express the transmitted symbols as ${x_A} = x_A^I + x_A^Qj$ and ${x_B} = x_B^I + x_B^Qj$. The received frequency-domain samples (our NCMA system is an OFDM system) \referred{FPNCPhycom12}\cite{FPNCPhycom12} on the two antennas of the AP are
\begin{align}
{y_{R1}} = {h_{A1}}{x_A} + {h_{B1}}{x_B} + {w_1}, \notag \\
{y_{R2}} = {h_{A2}}{x_A} + {h_{B2}}{x_B} + {w_2},
\label{equ:y1y2}
\end{align}
\noindent where ${h_{A1}}$ and ${h_{B1}}$ (${h_{A2}}$ and ${h_{B2}}$) are the uplink channel gains of nodes A and B associated with the first (second) antenna, respectively, and $w_1$, $w_2$ are additive white Gaussian noises (AWGN) with variances $\sigma^2_1$ and $\sigma^2_2$.
\begin{figure*}[t]
\centering
\includegraphics[width=0.72\textwidth]{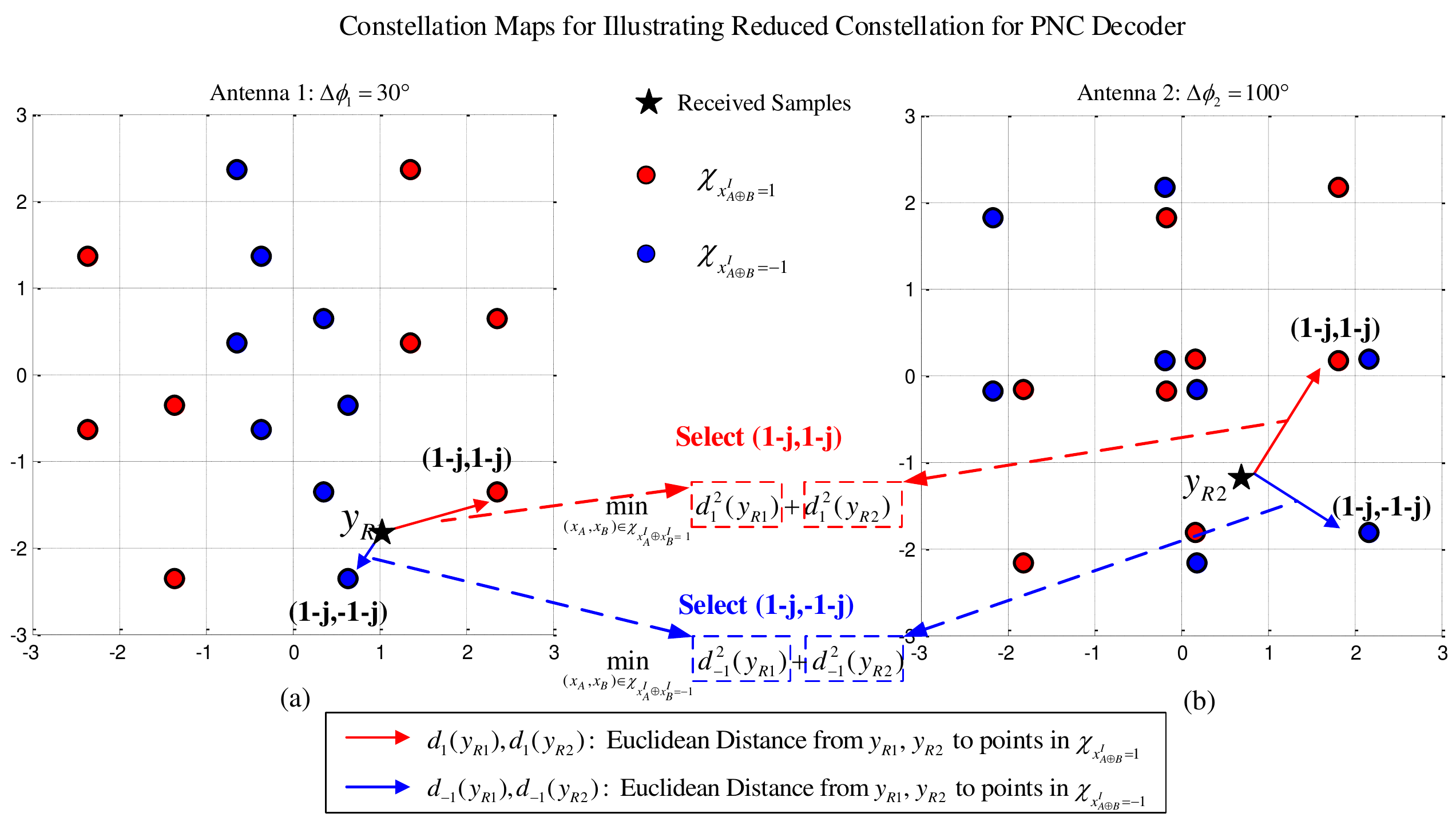}
\caption{Illustration of reduced-constellation scheme for PNC to obtain the soft information on $x_A^I \oplus x_B^I$: (a) Constellation map for antenna 1 with relative phase offset $\Delta {\phi _1} = 30^\circ $, (b) Constellation map for antenna 2 with relatvie phase offset $\Delta {\phi _2} = 100^\circ $. The black stars represent the received samples; red dots represent constellation points in ${\chi _{x_{A \oplus B}^I =  1}}$ and blue dots for ${\chi _{x_{A \oplus B}^I =  - 1}}$. Red and blue solid lines represent the Euclidean distances from received samples to the constellation points originated from the sets ${\chi _{x_{A \oplus B}^I =  1}}$ and ${\chi _{x_{A \oplus B}^I = -1}}$. We select the two constellation points by jointly considering two maps, i.e., the minimum sum of two Euclidean distances' squares (dashed lines). In this example, $(1-j,1-j)$ is selected in ${\chi _{x_{A \oplus B}^I =  1}}$, and $(1-j,-1-j)$ is selected in ${\chi _{x_{A \oplus B}^I = -1}}$. These two selected constellation points correspond to two different values of $x_A^I \oplus x_B^I$.}
\label{fig:pnc_mimo}
\end{figure*}

We next consider how to reduce the 16 constellation points of an overlapped QPSK joint symbol to the log-likelihood ratios of two binary XOR bits. Doing so reduces the complexity of the decoder design. We consider the in-phase XOR bit here; a similar procedure applies to the quadrature XOR bit. Define the in-phase component's LLR of packet A's QPSK symbol (i.e., $x_A$) as $\log ({P_A^I}/{Q_A^I})$, where $P_A^I$ and $Q_A^I$ are the probabilities of the in-phase component of $x_A$ being 1 and -1, respectively. Similarly, for $LLR(x_A^I \oplus x_B^I)$,  $P_{A \oplus B}^I$ and $Q_{A \oplus B}^I$ are the probabilities corresponding to $x_A^I \oplus x_B^I = 1$ and $x_A^I \oplus x_B^I = -1$. We have
\begin{align}
LLR(x_A^I \oplus x_B^I) &= \log P_{A \oplus B}^I - \log Q_{A \oplus B}^I \notag\\
&= \log \Pr (x_A^I \oplus x_B^I = 1|{y_{R1}},{y_{R2}}) \notag\\
&~~- \log \Pr (x_A^I \oplus x_B^I =  - 1|{y_{R1}},{y_{R2}}).
\label{equ:llrpnc}
\end{align}

Out of the 16 constellation points associated with the symbol pair $({x_A},{x_B})$, eight correspond to $x_A^I \oplus x_B^I = 1$ (the red dots in Fig. \ref{fig:pnc_mimo}), and eight correspond to $x_A^I \oplus x_B^I = -1$ (the blue dots in Fig. \ref{fig:pnc_mimo}). Let ${\chi _{x_{A \oplus B}^I = 1}}$ denote the set of symbol pairs $({x_A},{x_B})$  that satisfy $x_A^I \oplus x_B^I = 1$\footnote{The set ${\chi _{x_{A \oplus B}^I = 1}}$ contains eight constellation points originated from the symbol pair $({x_A},{x_B})$, namely, $(1+j, 1+j), (1+j, 1-j), (-1+j, -1+j), (-1+j,-1-j), (1-j, 1+j), (1-j, 1-j), (-1-j, -1+j)$ and $(-1-j, -1-j)$.}.  We can express $P_{A \oplus B}^I$ as
\begin{align}
&P_{A \oplus B}^I = \Pr (x_A^I \oplus x_B^I = 1|{y_{R1}},{y_{R2}})\notag \\
&\propto \sum\limits_{({x_A},{x_B}) \in {\chi _{x_{A \oplus B}^I = 1}}} {\exp \{  - \frac{{|{y_{R1}} - {h_{A1}}{x_A} - {h_{B1}}{x_B}{|^2}}}{{\sigma _1^2}}} \} \notag \\
& ~~~~\cdot { \exp\{ } - \frac{{|{y_{R2}} - {h_{A2}}{x_A} - {h_{B2}}{x_B}{|^2}}}{{\sigma _2^2}}{\rm{\} }}.
\label{equ:pnc_p1}
\end{align}

We compute $Q_{A \oplus B}^I$ in a similar way based on the set ${\chi _{x_{A \oplus B}^I =  - 1}}$. After that, we substitute $P_{A \oplus B}^I$ and $Q_{A \oplus B}^I$ into the LLR expression of (\ref{equ:llrpnc}) .
Fig. \ref{fig:pnc_mimo} plots the constellation maps of the two antennas with the same uplink channel-gain magnitude but with relative phase offsets $\Delta {\phi _1} = 30^\circ $ and $\Delta {\phi _2} = 100^\circ $ on antennas 1 and 2. Constellation points of sets ${\chi _{x_{A \oplus B}^I = 1}}$ and ${\chi _{x_{A \oplus B}^I = -1}}$ are marked by red and blue dots, respectively. In MIMO-NCMA, for further simplification, when computing $LLR(x_A^I \oplus x_B^I)$, we first reduce the number of constellation points from 16 to 2, i.e., choose only one constellation point in  ${\chi _{x_{A \oplus B}^I = 1}}$ and one in ${\chi _{x_{A \oplus B}^I = -1}}$ (see the red and blue arrows of Fig. \ref{fig:pnc_mimo}). The two selected constellation points correspond to the most likely points representing two different XOR values of $x_A^I \oplus x_B^I$. After that, we compute the LLR based on the two selected constellation points (see the dashed blocks between the two figures). We refer to this demodulation procedure as reduced-constellation demodulation.

\noindent \underline{\textbf{Reduced-constellation Demodulation for Two Antennas}}

We assume the noise variances $\sigma _1^2$ and $\sigma _2^2$ are the same, $\sigma _1^2 = \sigma _2^2 = \sigma^2$. Note that, in real wireless systems, $\sigma _1^2$ and $\sigma _2^2$ may not be equal; however, our derivations below can be easily generalized to deal with the case $\sigma _1^2 \ne \sigma _2^2$. We adopt the log-max approximation, $\log (\sum {_ie} xp({z_i})) \approx {\max _i}{z_i}$ \referred{FPNCPhycom12}\cite{FPNCPhycom12}, to simplify the LLR calculation. For example, $\log P_{A \oplus B}^I$ can be expressed as (\ref{equ:logpnc_p1}), where $d_1({y_{R1}}) = |{y_{R1}} - {h_{A1}}{x_A} - {h_{B1}}{x_B}|$ and $d_1({y_{R2}}) = |{y_{R2}} - {h_{A2}}{x_A} - {h_{B2}}{x_B}|$ are the Euclidean distances from ${y_{R1}}$ and ${y_{R2}}$ to the constellation point $({x_A},{x_B})$ in ${\chi _{x_{A \oplus B}^I = 1}}$. The physical meaning of (\ref{equ:logpnc_p1}) can be understood to be selecting one point with the minimum $d_1^2({y_{R1}}) + d_1^2({y_{R2}})$ value among all symbol pairs in set ${\chi _{x_{A \oplus B}^I = 1}}$.

Similarly, define $d_{-1}({y_{R1}})$ and $d_{-1}({y_{R2}})$  as the Euclidean distance from ${y_{R1}}$ and ${y_{R2}}$ to points in ${\chi _{x_{A \oplus B}^I = -1}}$, respectively. In Fig. \ref{fig:pnc_mimo}, $(1-j,1-j)$ is selected among the points in ${\chi _{x_{A \oplus B}^I = 1}}$, and $(1-j,-1-j)$ is selected among the points in ${\chi _{x_{A \oplus B}^I = -1}}$ to represent the cases of   $x_A^I \oplus x_B^I = 1$ and  $x_A^I \oplus x_B^I = -1$, respectively. The approximation of $LLR(x_A^I \oplus x_B^I)$ is
\begin{align}
LLR(&x_A^I \oplus x_B^I) \approx {\min _{({x_A},{x_B}) \in {\chi _{x_{A \oplus B}^I = 1}}}}\{ d_1^2({y_{R1}}) + d_1^2({y_{R2}})\} \notag \\
&- {\min _{({x_A},{x_B}) \in {\chi _{x_{A \oplus B}^I =  - 1}}}}\{ d_{ - 1}^2({y_{R1}}) + d_{ - 1}^2({y_{R2}})\}.
\label{equ:llr_final}
\end{align}

\newcounter{mytempeqncnt}
\begin{figure*}[!t]
\setcounter{mytempeqncnt}{\value{equation}}
\small
\begin{align}
\log P_{A \oplus B}^I &\propto {\max _{({x_A},{x_B}) \in {\chi _{x_{A \oplus B}^I = 1}}}}\{  - |{y_{R1}} - {h_{A1}}{x_A} - {h_{B1}}{x_B}{|^2} - |{y_{R2}} - {h_{A2}}{x_A} - {h_{B2}}{x_B}{|^2}{\rm{\} }} \notag \\
&{\rm{          }} \propto {\min _{({x_A},{x_B}) \in {\chi _{x_{A \oplus B}^I = 1}}}}\{ d_1^2({y_{R1}}) + d_1^2({y_{R2}})\} .
\label{equ:logpnc_p1}
\end{align}
\hrulefill
\end{figure*}

The QPSK demodulation from $x_A^I[k] \oplus x_B^I[k]$ to ${v_A}[2k - 1] \oplus {v_B}[2k - 1]$ is a one-to-one mapping (see (\ref{equ:pnc_demo})), and the following LLR relationship holds:
\begin{align}
LLR({v_A}[2k - 1] &\oplus {v_B}[2k - 1])=  LLR(x_A^I[k] \oplus x_B^I[k]).
\label{equ:llr_demod}
\end{align}

Similarly, even input bits' $LLR({v_A}[2k] \oplus {v_B}[2k])$ is $LLR(x_A^Q[k] \oplus x_B^Q[k])$.

\begin{figure*}[t]
\centering
\includegraphics[width=0.71\textwidth]{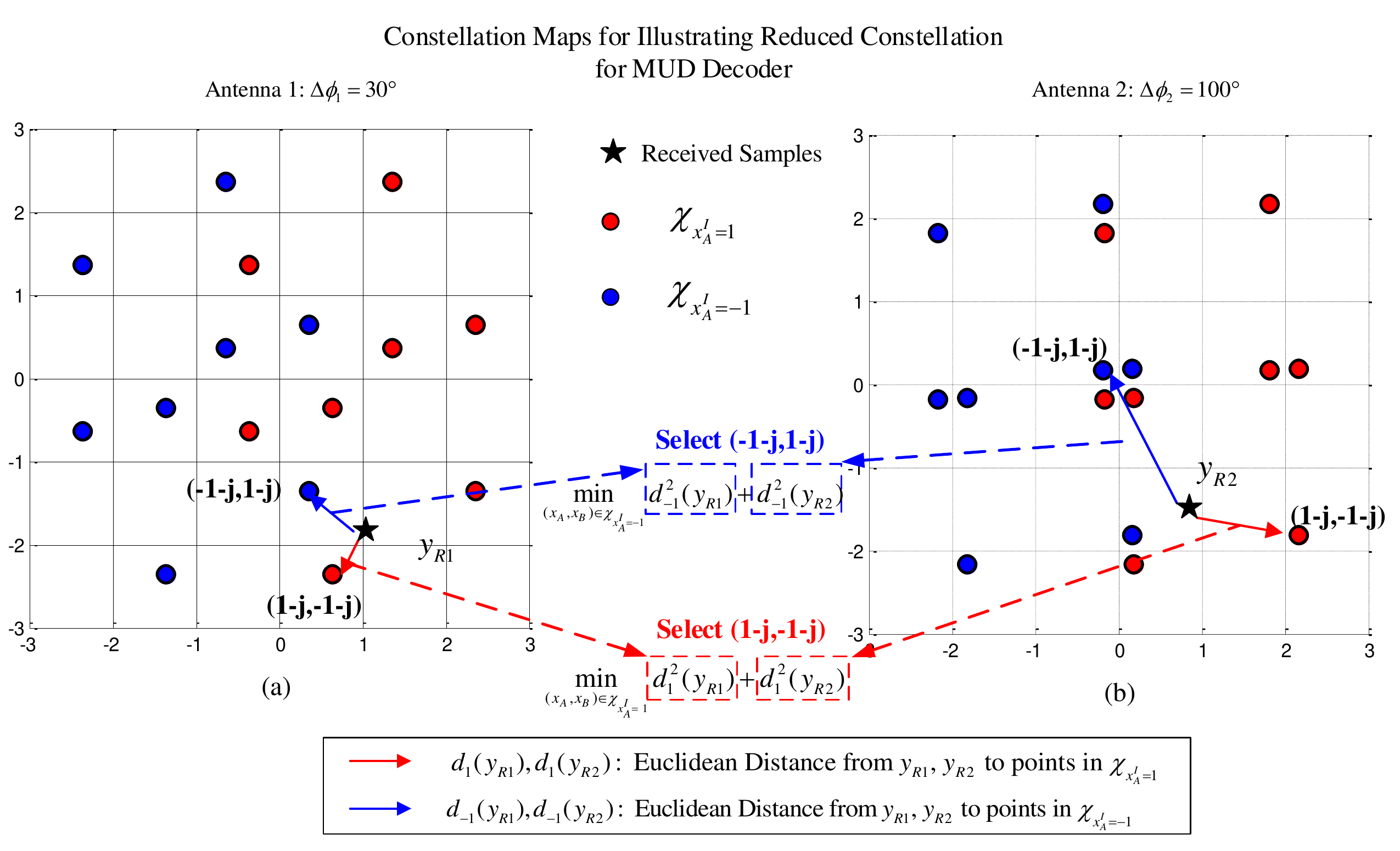}
\caption{Illustration of the reduced-constellation scheme for MUD to obtain soft information on $x_A^I$: Red and blue solid lines represent the Euclidean distances from received samples to the constellation points originated from the sets ${\chi _{x_A^I = 1}}$ and ${\chi _{x_A^I = -1}}$. In this example, $(1-j,-1-j)$ is selected in ${\chi _{x_A^I = 1}}$, and $(-1-j,1-j$) is selected in ${\chi _{x_A^I = -1}}$. These two selected constellation points correspond to two different values of  $x_A^I$.}
\label{fig:mud_mimo}
\end{figure*}

\subsection{MUD Decoder (MUD-CD)} \label{sec:MIMO_NCMA2}
The MUD decoder for MIMO-NCMA follows the same reduced-constellation principle as that of the PNC decoder, with the difference that its target is to obtain the individual soft information of ${\{ v_A[n]\} _{n = 1,2,...}}$ and ${\{ v_B[n]\} _{n = 1,2,...}}$ rather than their XOR. Without loss of generality, let us focus on the derivation of the soft information of packet A.  $LLR(x_A^I[k])$ and $LLR(x_A^Q[k])$ are the soft information for ${v_A}[2k - 1]$ and ${v_A}[2k]$, respectively. Based on the sets of $\chi _{x_A^I = 1}$\footnote{The set $\chi _{x_A^I = 1}$ also contains eight constellation points originated from the symbol pair $({x_A},{x_B})$, namely, $(1+j, 1+j), (1+j, 1-j), (1+j, -1+j), (1+j,-1-j), (1-j, 1+j), (1-j, 1-j), (1-j, -1+j).$ and $(1-j, -1-j)$}  and $\chi _{x_A^I = -1}$, and the same ``log-max approximation'' rule as in (\ref{equ:logpnc_p1}), we now have the approximation of $LLR(x_A^I)$:
\begin{align}
LLR(&x_A^I) \approx {\min _{({x_A},{x_B}) \in {\chi _{x_{A}^I = 1}}}}\{ d_1^2({y_{R1}}) + d_1^2({y_{R2}})\} \notag \\
&- {\min _{({x_A},{x_B}) \in {\chi _{x_{A}^I =  - 1}}}}\{ d_{ - 1}^2({y_{R1}}) + d_{ - 1}^2({y_{R2}})\}.
\label{equ:llrmud_final}
\end{align}


Fig. \ref{fig:mud_mimo} shows a reduced constellation example for MUD-CD using the same constellation map of Fig. \ref{fig:pnc_mimo}. We note that for the same constellation map and the same received samples, the soft outputs of XOR-CD decoder and MUD-CD decoder are in general different. In a real wireless system, since the constellation map changes from one sample to another sample, sometimes the MUD decoder's LLR may be higher, and sometimes the PNC's LLR. NCMA can best utilize and extract useful information out of the received samples, by jointly making use of PNC and MUD decoders.

\subsection{Simulation Results for Bit-level Decoding} \label{sec:MIMO_NCMA3}
We now take a look at the simulation results in Fig. \ref{fig:Ber_curve}(b), where two antennas and bit-level decoding are adopted. Fig. \ref{fig:Ber_curve}(b) plots the BER curves of QPSK PNC and MUD decoders with different $\Delta \phi_1$ and $\Delta \phi_2$, where $\Delta \phi_1$ and $\Delta \phi_2$ are the relative phase offsets at antenna 1 and antenna 2, respectively. When $\Delta {\phi _1}=\pi/2$, $\Delta {\phi _2}=0$, we can see that the BER performances of both XOR-CD and MUD-CD decoders are greatly improved compared with the single antenna case with a $\pi / 2$ phase offset. The joint use of MIMO and bit-level decoding solves the phase penalty issue for QPSK-modulated NCMA systems. 

However, for higher-order modulations, e.g., 16-QAM, bit-level decoding still incurs severe performance degradation, even with the use of MIMO. Fig. \ref{fig:Ber_curve}(c) plots the BER curves of 16-QAM bit-level PNC and MUD decoders. We can see from Fig. \ref{fig:Ber_curve}(c) that even with perfect phase synchronization (i.e., $\Delta {\phi _1}=\Delta {\phi_2}=0$), the 16-QAM bit-level XOR-CD decoder reaches a BER floor. Fortunately, as will be discussed in the next section, using symbol-level decoding improves the performance greatly (also see Fig. \ref{fig:Ber_curve}(c)).

\section{MIMO-NCMA: 16-QAM with Symbol-level Decoding} \label{sec:Symbol-level_NCMA}


This section studies \emph{symbol-level} decoding that does away bit-level decoding to avoid information loss in the demodulation process. To utilize the soft-symbol information, we investigate the combination of symbol-level PNC and MUD demodulators with a symbol-level Viterbi decoder, by generalizing the reduced-constellation algorithm. For simplicity and without loss of generality, we start with QPSK to explain the idea of symbol-level decoding and the potential performance degradation problem in bit-level decoders (although the performance degradation in QPSK is small when MIMO is used). After that, we extend the treatment from QPSK to 16-QAM.

\begin{figure*}[!t]
\small
\setcounter{mytempeqncnt}{\value{equation}}
\begin{align}
\log &\Pr (x_{A \oplus B}^{} = 1 + j|{y_{R1}},{y_{R2}}) \notag \\
&\propto \log \sum\limits_{({x_A},{x_B}) \in \chi _{x_{A \oplus B} = 1 + j}} {\exp \{  - \frac{{|{y_{R1}} - {h_{A1}}{x_A} - {h_{B1}}{x_B}{|^2}}}{{\sigma _1^2}}}\rm{\}}
{ {\exp\{ }} - \frac{{|{y_{R2}} - {h_{A2}}{x_A} - {h_{B2}}{x_B}{|^2}}}{{\sigma _2^2}}{\rm{\} }}.
\label{equ:llr_symbol}
\end{align}
\hrulefill
\end{figure*}

\subsection{QPSK Symbol-level NCMA Decoder}\label{sec:Symbol-level_NCMA1}
\begin{figure}
\centering
\includegraphics[width=0.5\textwidth]{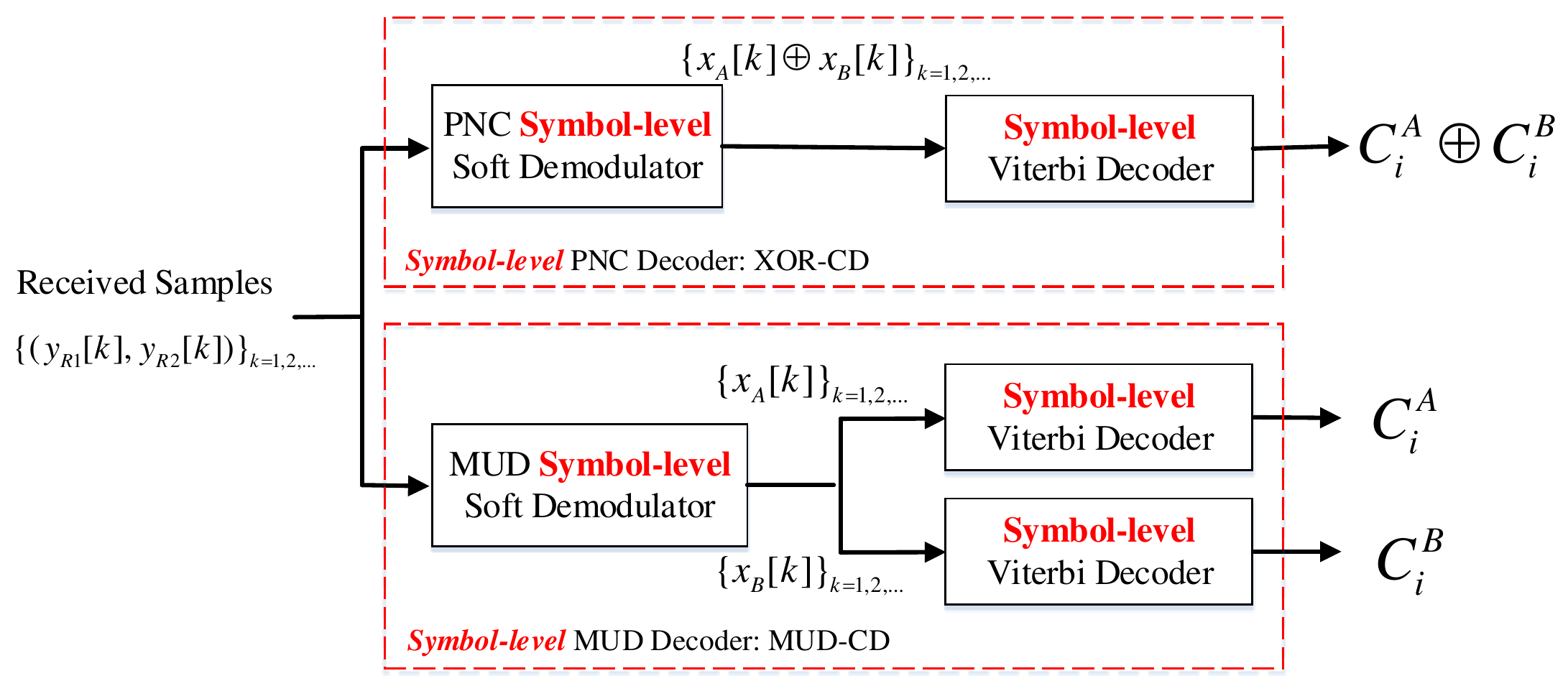}
\caption{Symbol-level NCMA decoder flow graph, in which the red characters highlight the difference with respect to the  bit-level decoder in Fig. \ref{fig:PHY-decoder}.}
\label{fig:PHY-symbol-decoder}
\end{figure}

The symbol-level NCMA decoder contains symbol-level demodulators and symbol-level Viterbi decoders, as shown in Fig. \ref{fig:PHY-symbol-decoder}. For example, consider the PNC decoder. Based on the received samples ${y_{R1}}$ and ${y_{R2}}$ from the two antennas in (\ref{equ:y1y2}), the QPSK symbol-level PNC demodulator generalizes the reduced-constellation demodulation scheme by reducing the 16 constellation points to 4 constellation points, i.e., we reduce to a whole QPSK XOR symbol rather than two bit-wise XORs. The soft information for $x_{A \oplus B} = 1 + j$, for example, is expressed as (\ref{equ:llr_symbol}), where ${\chi _{x_{A \oplus B} = 1 + j}}$ denotes the set of symbol pair $({x_A},{x_B})$ satisfying $x_{A \oplus B} = 1 + j$. We further express (\ref{equ:llr_symbol}) using the ``log-max" approximation:
\begin{align}
\log &\Pr (x_{A \oplus B}= 1 + j |{y_{R1}},{y_{R2}}) \notag \\
&\propto {\min _{({x_A},{x_B}) \in \chi _{x_{A \oplus B} = 1 + j}}}\{ d_{1 + j}^2({y_{R1}}) + d_{1 + j}^2({y_{R2}})\},
\label{equ:llr_symbol_final}
\end{align}

\noindent
where $d_{1 + j}({y_{R1}}) = |{y_{R1}} - {h_{A1}}{x_A} - {h_{B1}}{x_B}|$ ($d_{1 + j}({y_{R2}}) = |{y_{R2}} - {h_{A2}}{x_A} - {h_{B2}}{x_B}|$) is the Euclidean distance between ${y_{R1}}$ $({y_{R2}})$ and the constellation point $({x_A},{x_B})$ satisfying $x_{A \oplus B}=1+j$. We can compute $\log \Pr (x_{A \oplus B}=  - 1 + j |{y_{R1}},{y_{R2}})$,  $\log \Pr (x_{A \oplus B}=  1 - j |{y_{R1}},{y_{R2}})$ and $\log \Pr (x_{A \oplus B}=  - 1 - j |{y_{R1}},{y_{R2}})$ in a similar manner. The above describes the symbol-level demodulation process to obtain the soft XOR information for each QPSK symbol. This soft information is then fed to a symbol-level Viterbi decoder to decoder the original XOR packet $C_i^A \oplus C_i^B$.

\subsection{Performance Degradation Problem in Bit-level Decoders}\label{sec:Symbol-level_NCMA2}

To motivate the use of the symbol-level decoder, we now elaborate the performance degradation problem associated with the bit-level PNC decoder assuming QPSK. For simplicity in our illustration, let us assume that the relative phase offsets between nodes A and B are $\pi/2$ for both antennas, and that the amplitudes of all the channel gains are 1. Furthermore, let us first neglect the noise and consider the reception of two specific noise-free constellation points   $y[i] = 2$ and $y[i + 1] = 2 + 2j$  at the AP. Note that given the same relative offset $\pi/2$, the received signals at the two antennas are the same. Thus, we consider the signal on one of the antennas here for our illustration. It is easy to see that $y[i] = 2$ may result from two symbol pairs $(1-j, 1-j)$ and $(1+j, -1-j)$ with equal probability, and will be mapped into $1+j$ and $-1-j$ under XOR mapping with probability 1/2, as follows:
\begin{align}
&\Pr (x_{A \oplus B}[i] = 1 + j |y[i]) =\Pr (x_{A \oplus B}[i] =  - 1 - j |y[i]) \notag \\
=&1/2.
\label{equ:probability_1}
\end{align}

\noindent
On the other hand,  $y[i + 1]$ can only result from $(1+j, 1-j)$ and can be mapped to a unique XOR $1-j$. 

By contrast, note that the bit-level PNC demodulator splits a QPSK symbol into an in-phase bit and a quadrature bit. With respect to the same sample $y[i] = 2$, the bit-level PNC demodulator computes the probability of each bit independently as follows,
\begin{align}
\Pr (x_{A \oplus B}^I[i] = 1|y[i]) = \Pr (x_{A \oplus B}^I[i] =  - 1|y[i]) = 1/2, \notag \\
\Pr (x_{A \oplus B}^Q[i] = 1|y[i]) = \Pr (x_{A \oplus B}^Q[i] =  - 1|y[i]) = 1/2,
\label{equ:probability_2}
\end{align}
which is equivalent to
\begin{align}
&\Pr (x_{A \oplus B}[i] = 1 + j|y[i]) = \Pr (x_{A \oplus B}[i] =  - 1 - j|y[i])  \notag \\
= &\Pr (x_{A \oplus B}[i] = 1 - j|y[i]) = \Pr (x_{A \oplus B}[i] =  - 1 + j|y[i])  \notag \\
=& 1/4.
\label{equ:probability_3}
\end{align}

\begin{figure*}[t]
\centering
\includegraphics[width=0.78\textwidth]{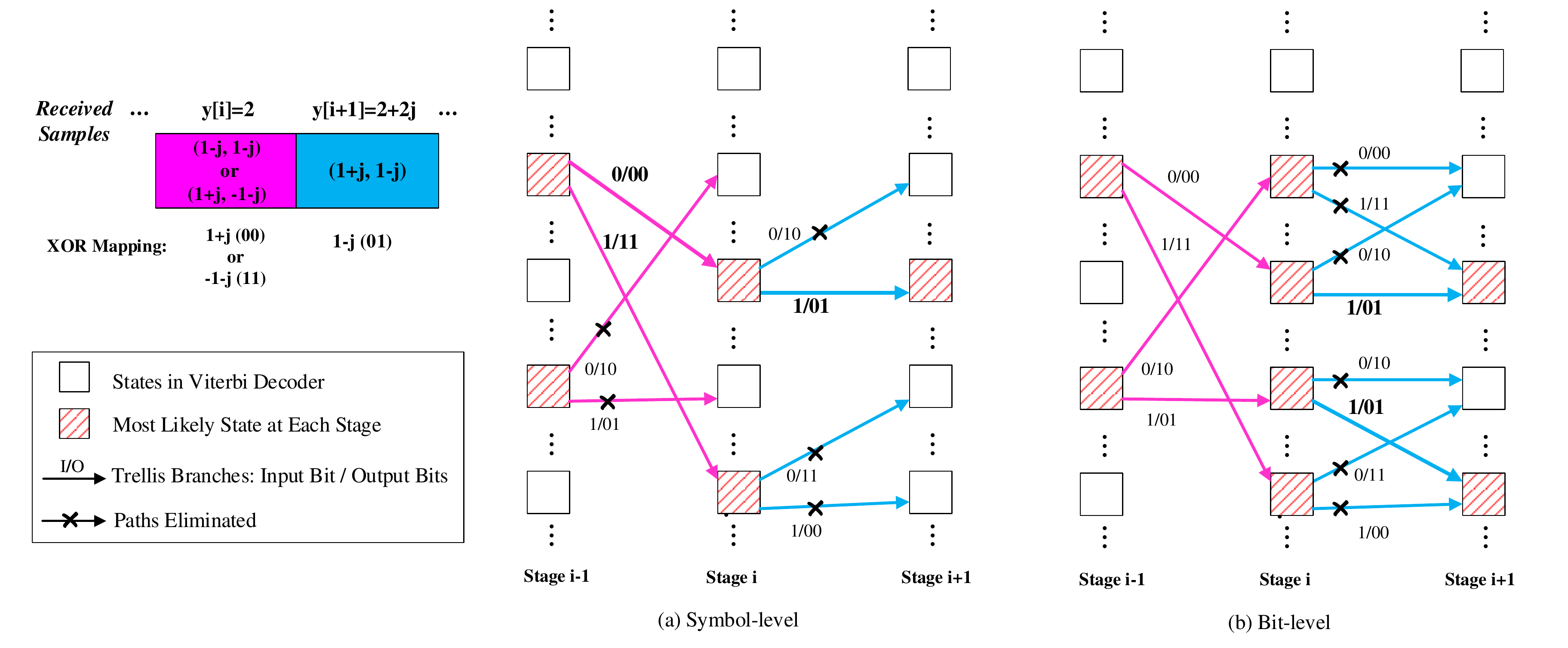}
\caption{Examples of trellis diagrams for Viterbi with two consecutive constellation points, $y[i] = 2$ followed by $y[i + 1] = 2 + 2j$: (a) a symbol-level decoder; (b) a bit-level decoder.}
\label{fig:symbol-level-example}
\end{figure*}

Comparing (\ref{equ:probability_3}) with (\ref{equ:probability_1}), it is obvious that the bit-level demodulator introduces ambiguity, which may cause performance degradation in the Viterbi decoding process. In particular, given that the four possible QPSK XORs are equally likely, there is no information contained in (\ref{equ:probability_3}) (i.e., this amounts to feeding no information to the Viterbi decoder).

The above explains the loss of information in the demodulation process. We now look at its impact on the decoding process. In particular, we examine the difference between the symbol-level and bit-level Viterbi decoder, and let us look at an example as shown in Fig. \ref{fig:symbol-level-example}. Without loss of generality, let us assume that at stage $i-1$ of their trellises, both the bit-level and symbol-level Viterbi decoders have the same path metric (i.e., the cumulated Euclidean distances), and only two states have much higher probability than other states so that we can look at these two states only for future decoding process. When $y[i] = 2$ is received, the symbol-level PNC demodulator computes the probability of a whole QPSK symbol as in (\ref{equ:probability_1}), which is then fed into the symbol-level Viterbi decoder. In Fig. \ref{fig:symbol-level-example}(a), the branches that output ``00" and ``11" are selected as survival paths from stage $i-1$ to stage $i$, and the number of possible states remains two. However, in Fig. \ref{fig:symbol-level-example}(b), where the trellis of the bit-level decoder is shown, all branches from stage $i-1$ to stage $i$ have equal branch metric due to (\ref{equ:probability_3}), and the number of possible states is double. With respect to the same point $y[i + 1]$ which is mapped to the unique XOR symbol $1-j$, it helps eliminate one ambiguity state by tracing back, and only one possible state is left in Fig. \ref{fig:symbol-level-example}(a). However, if the bit-level decoder is used in Fig. \ref{fig:symbol-level-example}(b), $y[i + 1]$ cannot reduce the number of possible states to one at stage $i+1$ because the number of possible states is doubled at stage $i$.

\begin{figure*}[t]
\normalsize
\small
\setcounter{mytempeqncnt}{\value{equation}}
\begin{align}
\log \Pr (x_{A \oplus B} = 3 + 3j|{y_{R1}},{y_{R2}}) &\propto {\max _{({x_A},{x_B}) \in \chi _{x_{A \oplus B}= 3 + 3j}}}\{  - |{y_{R1}} - {h_{A1}}{x_A} - {h_{B1}}{x_B}{|^2} - |{y_{R2}} - {h_{A2}}{x_A} - {h_{B2}}{x_B}{|^2}{\rm{\} }} \notag\\
&\propto {\min _{({x_A},{x_B}) \in \chi _{x_{A \oplus B} = 3 + 3j}}}\{ d_{3 + 3j}^2({y_{R1}}) + d_{3 + 3j}^2({y_{R2}})\}.
\label{equ:llr_16qam}
\end{align}
\hrulefill
\end{figure*}

\subsection{Higher-order Modulations beyond QPSK}\label{sec:Symbol-level_NCMA3}
In this subsection, we consider higher-order modulations beyond QPSK, and use 16-QAM as an example. As with QPSK, the four bits in an overlapped 16-QAM symbol are also correlated because of Gray mapping: their inter-bit dependency and correlation are more complicated than those in QPSK. Therefore, the information loss problem in 16-QAM bit-level decoders becomes more severe, and symbol-level decoding is needed. For 16-QAM, the same reduced-constellation techniques used in QPSK can be applied. In the following, we present the symbol-level PNC decoder design for 16-QAM as an example.

The 16-QAM symbol-level PNC demodulator tries to reduce the 256 constellation points to 16 constellation points for demodulation. The soft symbol-level information will be fed into the 16-QAM Viterbi decoder to decode the XORed packet. The soft information of each symbol can be obtained using the equation in (\ref{equ:llr_16qam}), where we choose symbol $3+3j$ as an example. The soft symbol-level information will be fed into a 16-QAM Viterbi decoder to decode the XOR packet.

\begin{figure*}[t]
\centering
\includegraphics[width=0.77\textwidth]{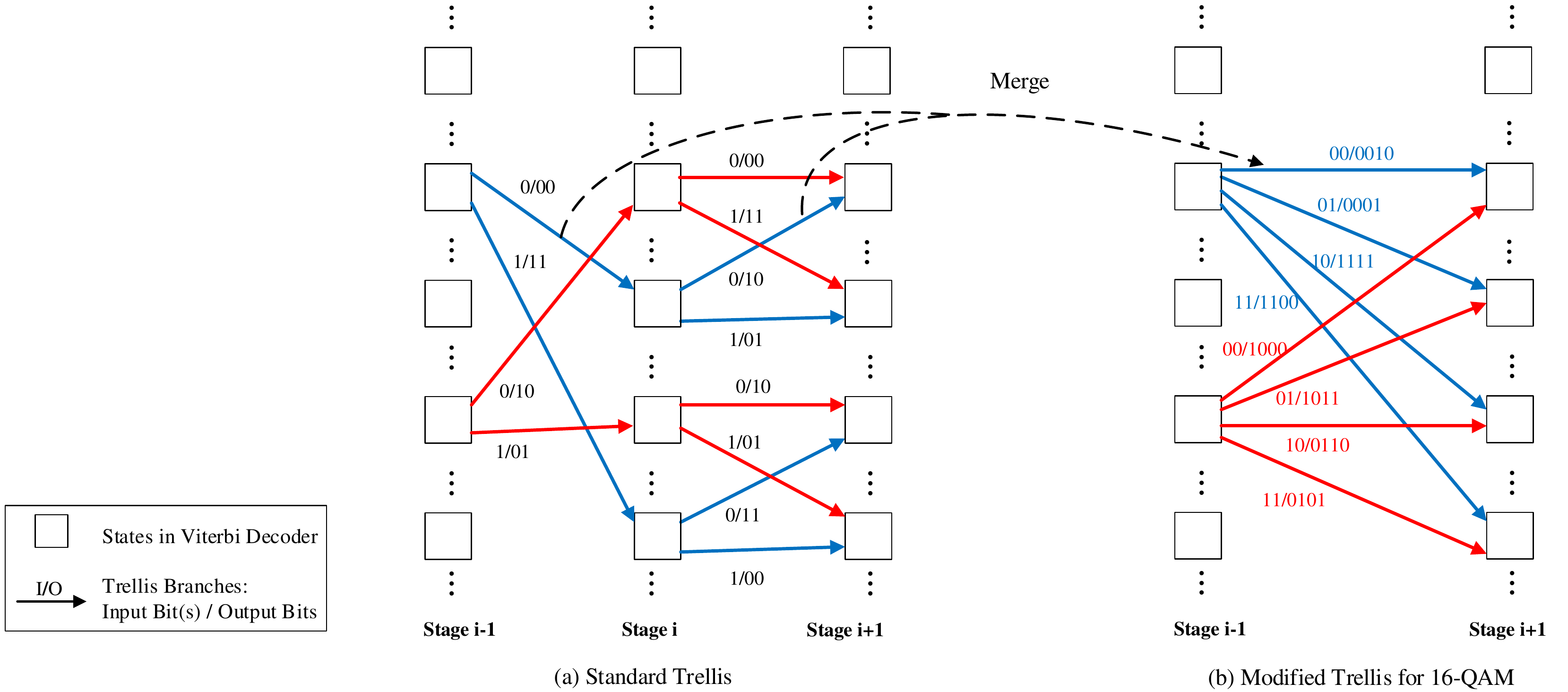}
\caption{Trellis diagrams for symbol-level Viterbi decoders using (a) QPSK; (b) 16-QAM. Every two adjacent state transitions in (a) are merged into one state transition in (b).}
\label{fig:trellis_16qam}
\end{figure*}

We remark that 16-QAM's demodulation complexity increases significantly compared with that of QPSK. This is due to the increase of the constellation size from 16 points to 256 points (QPSK to 16-QAM), and the 16-QAM demodulator has to compute 256 Euclidean distances to select points under the reduced-constellation principle. Considering that the joint symbol to XOR mapping is irregular because different nodes have different channel gains (i.e., the constellation map is not in a nice symmetric form because receiver-side equalization of the two channel gains simultaneously is impossible for PNC), exhaustive search may be required. Fortunately, unlike for PNC, the decision regions for the MUD can be reshaped into a regular 16-QAM shape \referred{dot11std13}\cite{dot11std13} using successive interference cancellation (SIC) and equalization. Based on this property, we put forth a reduced complexity 16-QAM NCMA demodulator in Appendix \ref{sec:appendix_MUD}. We further show that the BER performance of the reduced complexity MUD decoder is exactly the same as that of the exhaustive-search MUD decoder. After that, we further extend the reduced complexity principle from MUD to PNC in Appendix \ref{sec:appendix_PNC}.

For the channel decoder, we remind the reader that each end node still adopts the IEEE 802.11 1/2 convolutional code \referred{dot11std13}\cite{dot11std13}, and every input bit generates two output bits in one state transition (see I/O in Fig. \ref{fig:trellis_16qam}(a)). Since one QPSK symbol has two bits, which matches the number of output bits in one state transition, the trellis diagram in Fig. \ref{fig:trellis_16qam}(a) can be directly used by the QPSK symbol-level Viterbi decoder.

However, each 16-QAM symbol contains four bits and the soft information of one 16-QAM symbol is related to two adjacent state transition outputs. To utilize the 16-QAM symbol information, we have to modify the trellis diagram of standard 1/2 convolutional codes. The modifications are shown in \ref{fig:trellis_16qam}(b). In particular, we merge two adjacent state transitions (of QPSK) into one state transition. For example, the two consecutive branches ``0/00'' and ``0/10'' in Fig. \ref{fig:trellis_16qam}(a) are merged to a new branch ``00/0010'' in \ref{fig:trellis_16qam}(b). Note that the number of states in the new trellis diagram remains to be the same as the original trellis diagram, and each state has four outgoing branches (namely, ``00'', ``01'', ``10'' and ``11''), and the total number of stages in the new trellis diagram is half that of the original trellis diagram.

The BER performance of the 16-QAM symbol-level PNC and MUD decoders are plotted in Fig. \ref{fig:Ber_curve}(c). We can see that when $\Delta {\phi _1}=\Delta {\phi_2}=0$, the 16-QAM symbol-level XOR-CD decoder improves the BER performance significantly, and the symbol-level MUD-CD decoder improves very little. In general, the symbol-level PNC decoder improves more than the symbol-level MUD decoder, relative to their bit-level counterparts, respectively (also see the experimental results in Section \ref{sec:Exp22}). For simplicity, we use QPSK as an example to explain this phenomenon in Appendix \ref{sec:appendix_reason}. When $\Delta {\phi _1}=\pi/2$ and $\Delta {\phi _2}=0$, the symbol-level MUD-CD decoder also outperforms its bit-level counterpart. This suggests that the combining of MIMO (to increase degrees of freedom) and symbol-level decoding (to avoid information loss) is a practical solution to tackle the subtleties in 16-QAM NCMA systems.  

In the next section, we will evaluate MIMO-NCMA decoders' performances through experiments on software-defined radio.  The experimental settings differ from the above simulation settings in two major ways: (1) in a real system, $\Delta {\phi _1}$ and $\Delta {\phi _2}$ may not be exactly 0, $\pi/4$ or $\pi/2$; (2) more importantly, in a broadband system such as OFDM, $\Delta {\phi _1}$ and $\Delta {\phi _2}$ may vary across different subcarriers (i.e., not all samples within a packet incur the same relative phase offsets).

In other words, the simulations only serve to highlight certain points, but do not reflect what actually happen in practice. The experimental study (e.g., PHY-layer decoding statistics and packet decoding ratios) serves to fill the gap not covered by the simulation study and the merit of MIMO-NCMA will need to be validated in a real wireless communications system.

\begin{figure*}[t]
\centering
\includegraphics[width=0.8\textwidth]{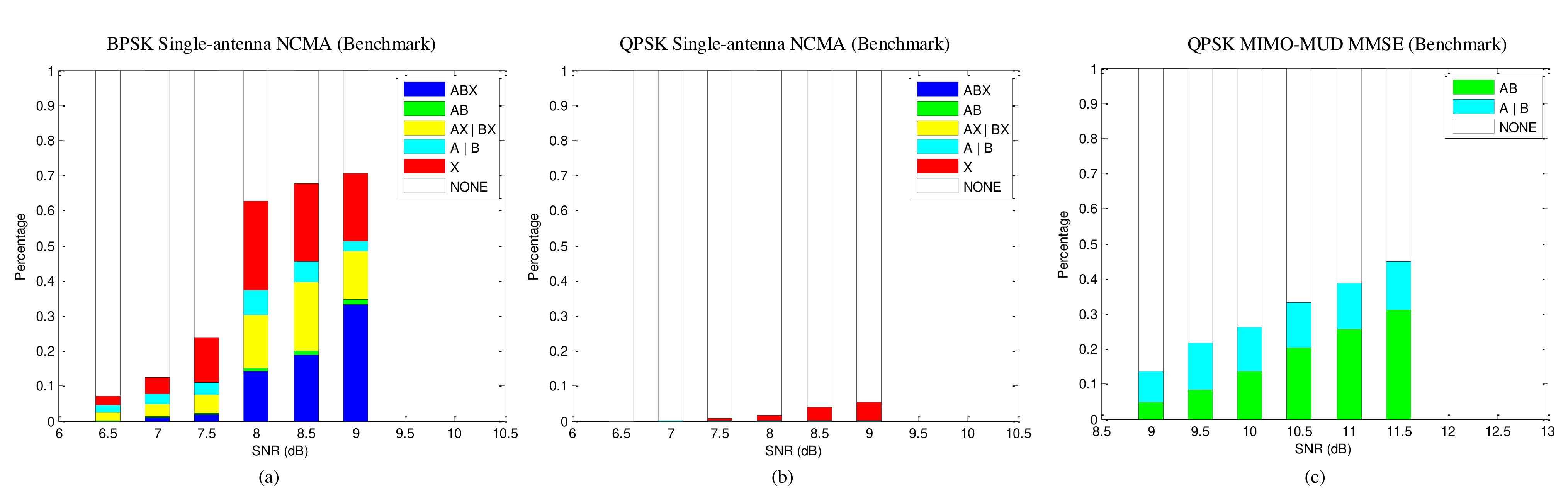}
\caption{PHY-layer packet statistics versus SNR for different benchmark decoding schemes with balanced received powers of two end nodes: (a) BPSK Single-NCMA; (b) QPSK Single-NCMA; (c) QPSK MIMO-MUD (MMSE).}
\label{fig:phy_stat_benchmark}
\end{figure*}

\section{Experimental Results} \label{sec:Exp}

For performance evaluation, we implemented MIMO-NCMA on software-defined radio. Section \ref{sec:Exp1} presents the implementation details and experimental setup and Section \ref{sec:Exp2} presents the experimental results.

\subsection{Implementation Details and Experiment Setup}\label{sec:Exp1}

The MIMO-NCMA system was built on the USRP hardware \referred{Ettus}\cite{Ettus} and the GNU Radio software with the UHD driver. We extended the single-antenna BPSK NCMA system in \referred{NCMA2}\cite{NCMA2} as follows:
\begin{itemize}\leftmargin=0in
\item [a)] We added one more antenna at the AP and changed the SISO system of \referred{NCMA2}\cite{NCMA2}  so that the AP can receive data from the two antennas. The end nodes still use one antenna.
\item [b)] We modified the transceiver design so as to support QPSK and 16-QAM in addition to BPSK.
\item [c)]We realized the QPSK and 16-QAM XOR-CD and MUD-CD decoders based on the reduced-constellation demodulation principle, incorporating symbol-level decoding.
\end{itemize}

For experimentation, we deployed three sets of USRP N210s with SBX daughterboards. Each MIMO-NCMA end node is one USRP connected to a PC through an Ethernet cable, and the MIMO-NCMA AP has two USRPs connected through one MIMO cable to behave like one node with two antennas. For the uplink channel, the AP sends beacon frames to trigger node A's and node B's simultaneous transmissions. Our experiments were carried out at 2.585GHz center frequency with 5MHz bandwidth.

To benchmark MIMO-NCMA, we consider three systems:
\begin{enumerate}\leftmargin=0in
\item \emph{Single-antenna NCMA system (Single-NCMA)} \\
This system is based on the previous single-antenna NCMA \referred{NCMA2}\cite{NCMA2}. In this system, all nodes have only one antenna each. In this system, all nodes have only one antenna each.We extend the system in \referred{NCMA2}\cite{NCMA2}  to support QPSK modulation in addition to BPSK. Both MUD decoder and PNC decoder are used. PHY-layer bridging and MAC-layer bridging are performed in the decoding process to increase the system throughput.

\item \emph{Distributed MIMO System (MIMO-MUD)} \\
This is a distributed MIMO-MUD system, where the receiver at the AP has two antennas and the transmitters at the two end nodes have only one antenna each. Conventional hard-input-hard-output ZF (zero-forcing) and MMSE (minimum mean square error) decoders are adopted \referred{tse2005fundamentals}\cite{tse2005fundamentals} for MUD decoding.

\item \emph{MIMO NCMA system (MIMO-NCMA)} \\
This is the high-order NCMA system proposed in this paper. Both XOR-CD and MUD-CD decoders are adopted. PHY-layer and MAC-layer bridgings are used
\end{enumerate}

\begin{figure}[!t]
\centering
\includegraphics[width=0.42\textwidth]{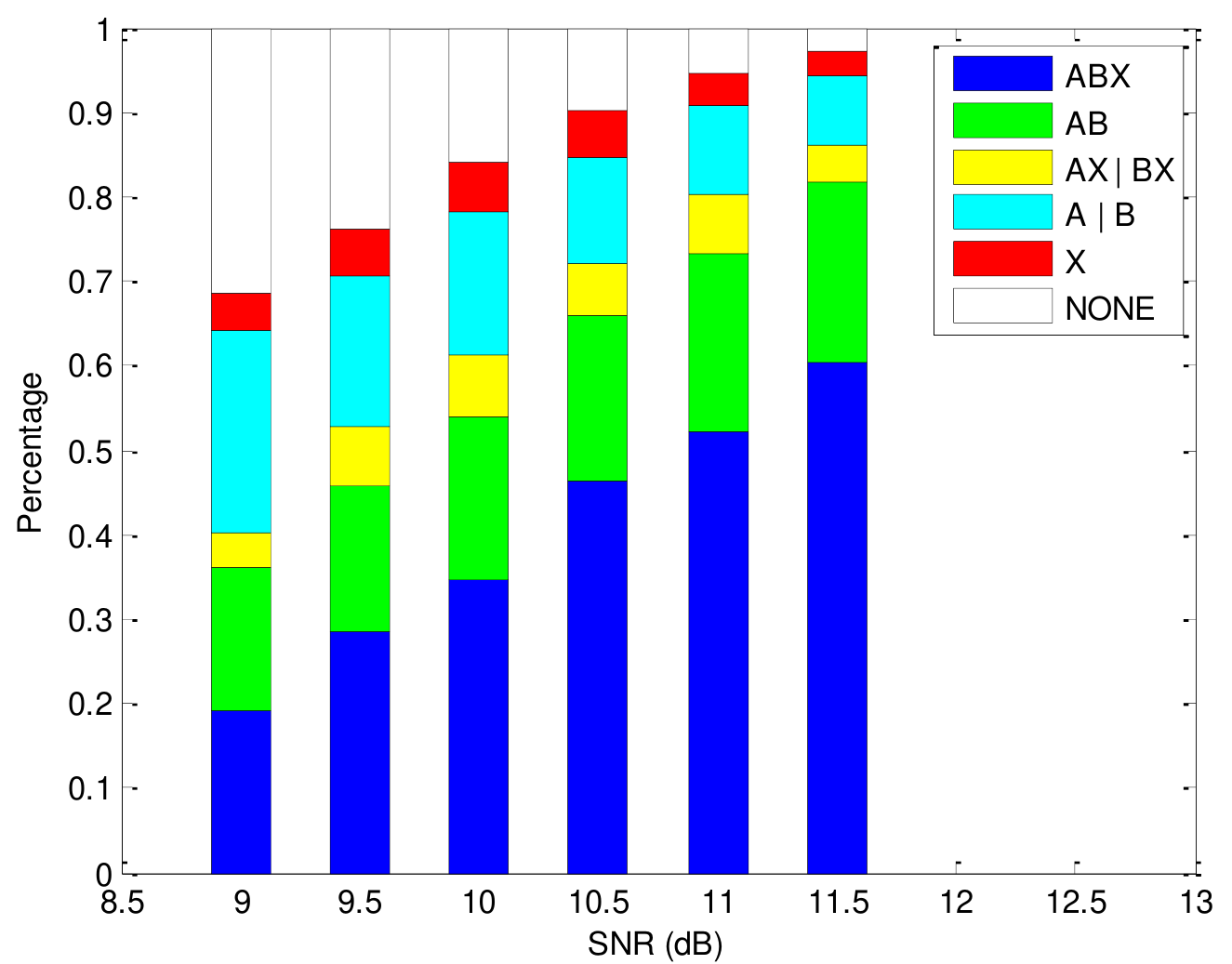}
\caption{MIMO-NCMA PHY-layer packet statistics comparison with the bit-level decoder using QPSK.}
\label{fig:phy_stat_mimoncma}
\end{figure}

\subsection{Experimental Results}\label{sec:Exp2}
We study both PHY-layer and MAC-layer performances: we first consider the PHY-layer packet decoding statistics of bit-level PNC and MUD decoders, and then we compare the decoding performance of the bit-level and symbol-level MIMO-NCMA decoder. After that, we evaluate the MAC-layer throughput.

\subsubsection{Bit-level Decoding Performance}\label{sec:Exp21}
We first study the bit-level decoding performance. We collected the QPSK PHY-layer decoding statistics of Single-NCMA, MIMO-MUD and MIMO-NCMA, and present the results in Fig. \ref{fig:phy_stat_benchmark} and Fig. \ref{fig:phy_stat_mimoncma} (all systems use bit-level decoding). There are eight possible decoding outcomes in each time slot (see Section \ref{sec:Overview2}) when PNC and MUD decoders are used jointly in Single-NCMA and MIMO-NCMA systems. We group some outcomes together as follows:
\begin{itemize}\leftmargin=0in
\item \texttt{NONE} = (iv)(b) (no packet decoded).
\item \texttt{X} = (iv)(a) (only XOR packet decoded).
\item \texttt{A|B} = (ii)(b) + (iii)(b) (either only packet $A$ or only packet $B$ decoded).
\item \texttt{AX|BX} = (ii)(a) + (iii)(a) (XOR packet plus either packet $A$ or packet $B$ decoded).
\item \texttt{AB} = (i)(b) (both packets $A$ and $B$ decoded; XOR packet not decoded).
\item \texttt{ABX} = (i)(a) (both packets $A$ and $B$ decoded; XOR packet decoded).
\end{itemize}

\begin{figure*}[t]
\centering
\includegraphics[width=0.75\textwidth]{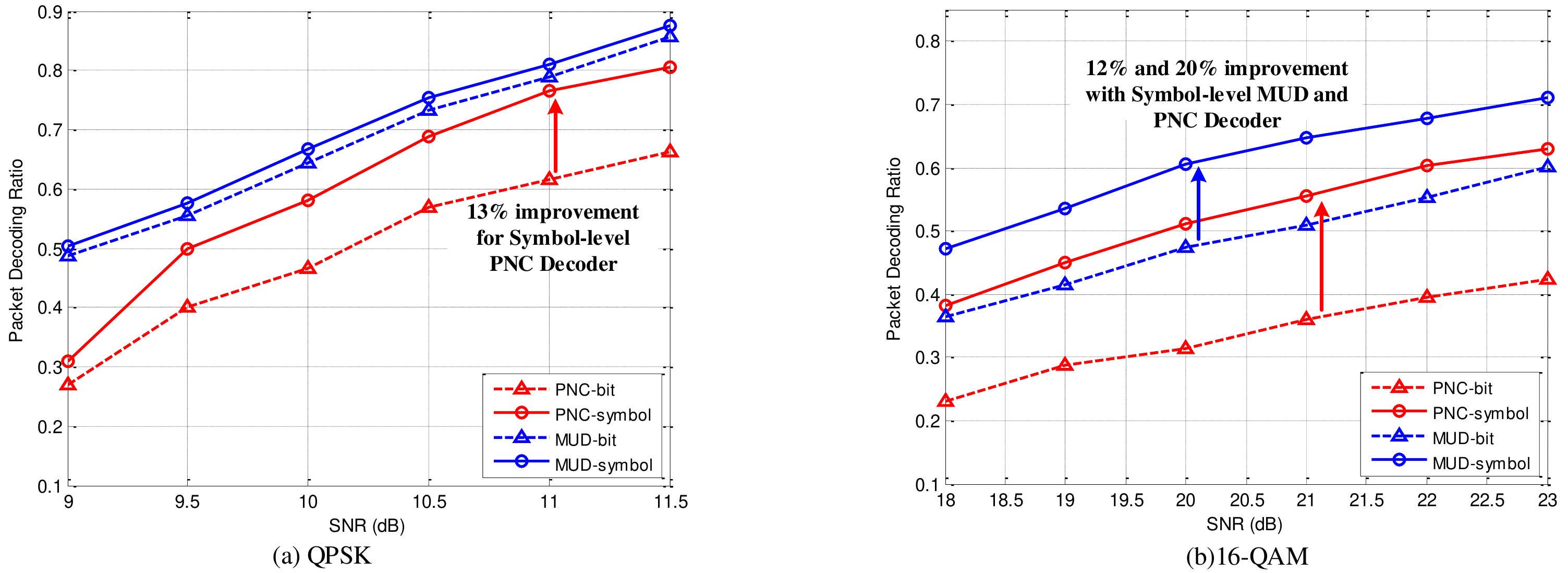}
\caption{Packet Decoding Ratio Comparison by using Bit-level and Symbol-level NCMA decoders with (a) QPSK and (b) 16-QAM.}
\label{fig:symbol_vs_bit}
\end{figure*}

We also plot the packet decoding ratios of PNC and MUD decoders versus SNR with (a) QPSK and (b) 16-QAM in Fig. \ref{fig:symbol_vs_bit}, including bit-level and symbol-level decoding, and let us focus on the bit-level decoding in this subsection (we define the PNC packet decoding ratio as the number of decoded PNC packets divided by the total number of time slots; the MUD packet decoding ratio is the number of decoded native packets divided by the total number of transmit packets from nodes A and B).

We performed controlled experiments for different received SNRs, and the received powers of signals from nodes A and B at the AP were adjusted to be approximately balanced (we remark that the powers of each pair could be slightly different due to channel fading, and the SNR presented here is the average SNR of all the received packets). We calculated the SNR using the method in \referred{HalperinSNR10}\cite{HalperinSNR10}, and varied the SNR values from 6.5 to 9 dB when the AP has single antenna. When the AP has two antennas, for QPSK, we varied the SNR from 9 to 11.5 dB since the effective receive power using the two antennas is almost doubled compared with single antenna case \referred{tse2005fundamentals}\cite{tse2005fundamentals}. For 16-QAM, we varied the SNR from 18 to 23 dB. For each SNR value, the AP sent 1,000 beacon frames to trigger simultaneous transmissions of two end nodes and each uplink packet has a 400-byte payload.

\noindent \textbf{\underline{Observation 1}: Single-NCMA fails to support QPSK}

Sections \ref{sec:SingleAntenna1} and \ref{sec:SingleAntenna2} have discussed the potential phase penalty associated with PNC and MUD decoders when QPSK is adopted in a single-antenna NCMA system. Our experimental results corroborate the theoretical and simulation analysis. The PHY-layer decoding statistics in Fig. \ref{fig:phy_stat_benchmark}(b) show that both PNC and MUD decoders cannot work well. Even at SNR=9dB (the working region for a point-to-point QPSK WLAN system\referred{HalperinSNR10}\cite{HalperinSNR10}), there are almost no decoded packets.

\noindent \textbf{\underline{Observation 2}: MIMO-NCMA works well for QPSK}

From Fig. \ref{fig:phy_stat_mimoncma}, we can see that the number of decoded packets for both bit-level PNC and MUD decoders increase drastically for MIMO-NCMA, compared with Single-NCMA of Fig. \ref{fig:phy_stat_benchmark}(b). At 9dB, around 70\% packets can be decoded correctly (either single packet or two packets), and at 11.5dB, the PER can be as low as less than 5\%. The performance of both PNC and MUD decoders improves with the additional antenna. Furthermore, we can see that the MUD decoder based on reduced-constellation principle has a better PER performance than the conventional MMSE decoder\footnote{Fig. \ref{fig:phy_stat_benchmark}(c) shows the PHY-layer statistics of the MMSE decoder. Although having two antennas at the AP improves the PER performance of single-antenna NCMA in Fig. \ref{fig:phy_stat_benchmark}(b), the performance of MMSE is still subpar (i.e., still less than 50\% packets are decoded correctly). } for QPSK (e.g., sum up the decoding outcomes ``ABX'' and `'AB'' in Fig. \ref{fig:phy_stat_mimoncma}, and then compare with ``AB'' in Fig. \ref{fig:phy_stat_benchmark}(c)), which is consistent with the BPSK results in \referred{NCMA1}\cite{NCMA1}.

\noindent \textbf{\underline{Observation 3}: MIMO-NCMA with Bit-level decoding does not perform well for 16-QAM}

In Fig. \ref{fig:symbol_vs_bit}(a), we can see that the QPSK bit-level MUD decoder has a packet decoding ratio over 80\% at SNR of 11.5dB. However, the bit-level decoder does not perform well when 16-QAM is adopted. Fig. \ref{fig:symbol_vs_bit}(b) shows that the packet decoding ratios for both 16-QAM bit-level PNC and MUD decoders are below 50\% at 20dB. We next describe results showing that symbol-level decoding can improve the performance of 16-QAM PNC and MUD decoders.



\subsubsection{Symbol-level Decoding Performance}\label{sec:Exp22}
As discussed in Section \ref{sec:Symbol-level_NCMA}, bit-level PNC and MUD demodulators lead to information loss when high-order modulations are used. In the following, we compare the performances of bit-level and symbol-level MIMO-NCMA decoders. From Fig. \ref{fig:symbol_vs_bit}, we can see that the symbol-level PNC decoder improves more than the symbol-level MUD decoder, relative to their bit-level counterparts, respectively. An intuitive explanation for the different improvements between symbol-level PNC and MUD decoders are referred to Appendix \ref{sec:appendix_reason}.


For QPSK MIMO-NCMA, bit-level decoding is already good and symbol-level decoding does not bring much improvement. The reason is that the bit-level MUD decoder is already very good in decoding both packets $A$ and $B$ in most time slots (e.g., in Fig. \ref{fig:phy_stat_mimoncma}, the success rate of decoding both packets $A$ and $B$ is over 70\% at 11dB). Fig. \ref{fig:symbol_vs_bit}(a) also shows that, for QPSK,  symbol-level MUD decoder has little improvement over bit-level MUD decoder.  Furthermore, because the MUD decoder can already obtain both packets $A$ and $B$ in most time slots, XOR packets obtained by the PNC decoder do not give much further improvement.  Although symbol-level PNC decoder can obtain more XOR packets than bit-leve PNC decoder, the extra XOR packets do not help when both packets $A$ and $B$ are already available. We will see in Section \ref{sec:Exp23} soon that QPSK MAC-layer throughput for the symbol-level decoder improves very little.

\noindent \textbf{\underline{Observation 4}: MIMO-NCMA with Symbol-level decoding improves 16-QAM performance significantly}

Unlike with QPSK, we find that with 16-QAM, the symbol-level decoders of both PNC and MUD outperform their bit-level counterparts substantially. From Fig. \ref{fig:symbol_vs_bit}(b) we can see that at SNR of 20dB, the packet decoding ratios for symbol-level PNC and MUD decoders increase by 20\% and 12\% over their corresponding bit-level decoders. This large improvement is attributed to the bit-level decoders losing too much information when mapping one 16-QAM symbol into four independent BPSK symbols. 

\begin{figure*}[t]
   \begin{minipage}{0.45\linewidth}
     \centering
     \includegraphics[width=0.78\textwidth]{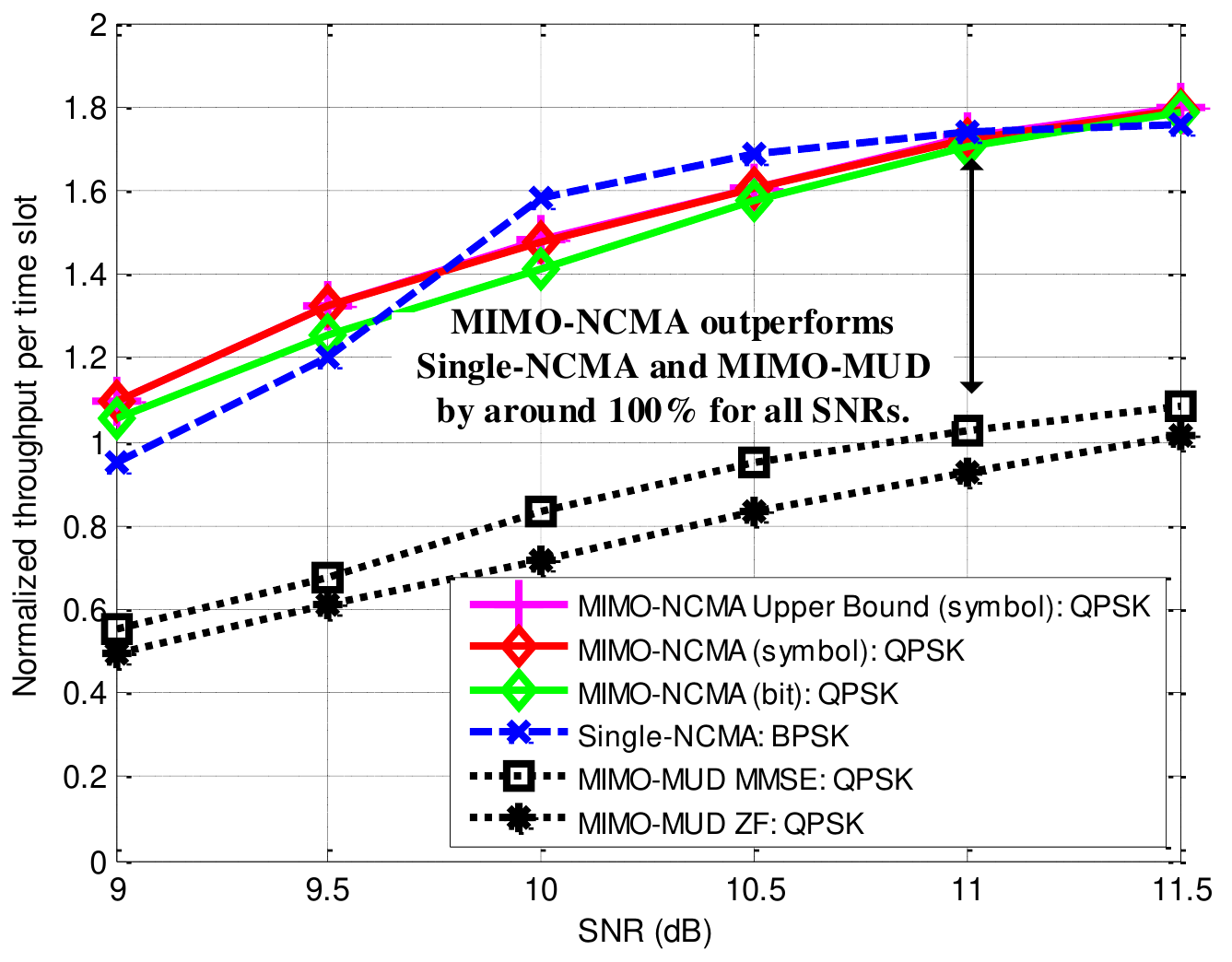}
\caption{MAC-layer performance: Throughput comparison of different schemes under different SNRs with RS code's constraint length ${L_A} = 1.5{L_B} = 24$. }
\label{fig:mac_throughput_qpsk}
   \end{minipage}
   \hfill
   \begin{minipage}{0.52\linewidth}
      \centering
\includegraphics[width=0.65\textwidth]{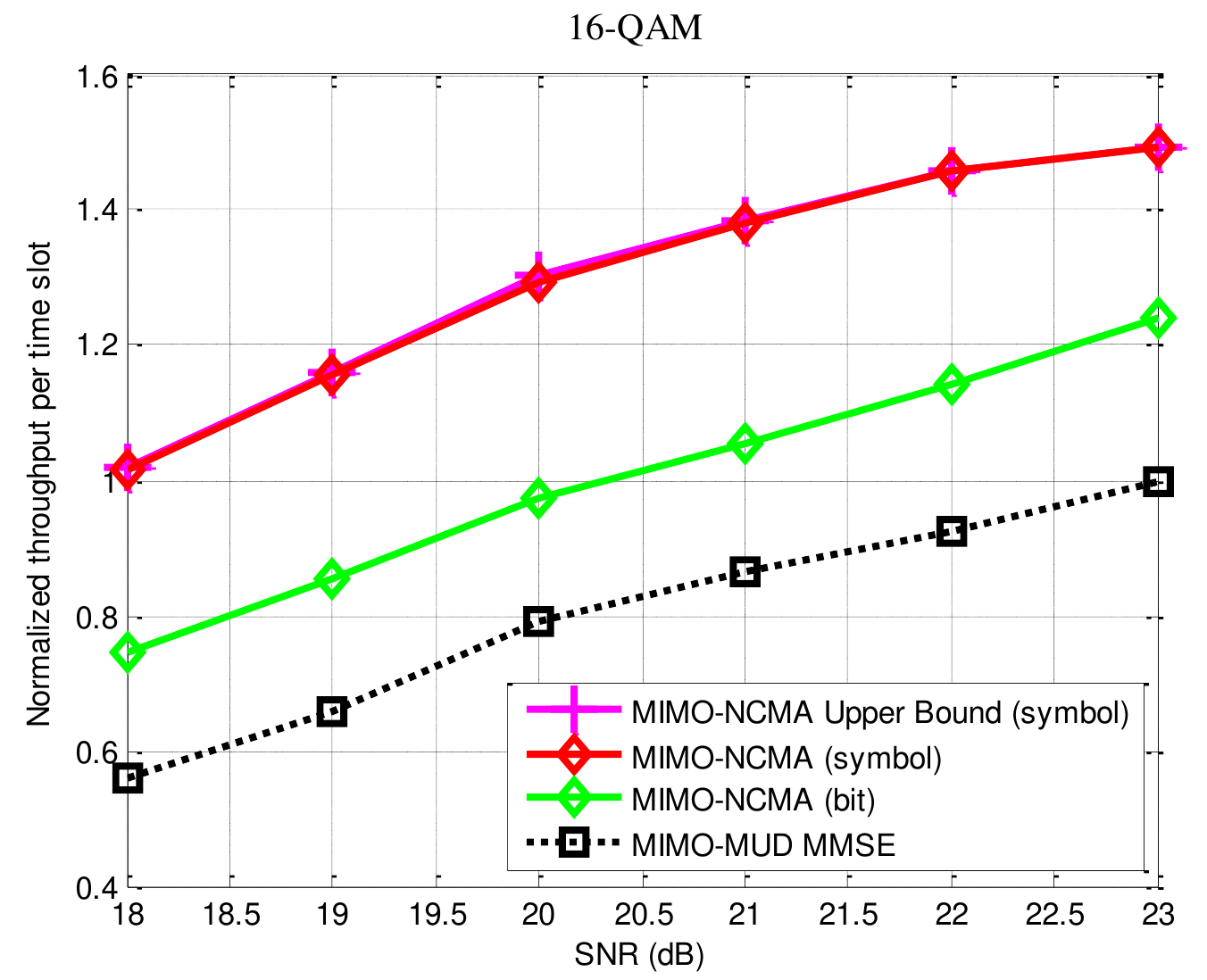}
\caption{MAC-layer performances of MIMO-NCMA with 16-QAM: Throughput comparison of different schemes under different SNRs with RS code's constraint length ${L_A} = 1.5{L_B} = 24$.}
\label{fig:mac_throughput_16QAM}
   \end{minipage}
\end{figure*}


\noindent \textbf{\underline{Observation 5}: Symbol-level NCMA decoder supports real-time operation}

In view of our final goal to design real-time NCMA decoders, a natural question here is that whether this performance gain is accompanied by increased decoding complexity. Next, we evaluate the real-time decoding speeds by measuring the processing time for the two main components in NCMA decoders: (1) the channel decoder, and (2) the demodulator. In Appendix \ref{sec:appendix_complexity}, we show that the complexities of the symbol-level and bit-level Viterbi decoders are the same for QPSK and 16-QAM modulations. But for 64-QAM (or even higher-order modulations), the symbol-level Viterbi decoder has a higher complexity\footnote{For the 64-QAM symbol-level Viterbi decoder, if a rate-1/2 convolutional code is used, we merge every three adjacent state transitions of the trellis into one state transition. Every state in the merged trellis has eight output branches.}. We implemented the symbol-level Viterbi decoder using GNU Radio \emph{gr-trellis} library \referred{GNURadio}\cite{GNURadio}. The measured decoding time confirms that the symbol-level and bit-level GNU Radio decoders are comparable in speed (e.g., around 2ms), which corroborates the theoretical complexity analysis (see Appendix \ref{sec:appendix_complexity} for details).

For demodulators, we can see from Appendix \ref{sec:appendix_complexity} that the processing time for QPSK MIMO-NCMA is small and comparable to the previous BPSK single-antenna case. However, the demodulation time for 16-QAM increases significantly due to the increase of constellation size, as discussed in Section \ref{sec:Symbol-level_NCMA3}. Fortunately, with our proposed reduced complexity demodulators in Appendices \ref{sec:appendix_MUD} and \ref{sec:appendix_PNC}, the demodulation time is greatly reduced after simplification (e.g., see Appendix \ref{sec:appendix_complexity} for the reduced demodulation times and Appendix \ref{sec:appendix_MUD} for BER performance comparisons).

\subsubsection{MAC-layer Throughput Performance}\label{sec:Exp23}
We now evaluate the overall NCMA throughput performance at the MAC layer. In NCMA, the PHY layer could decode one or two independent packets in one time slot (we treat the cases of ABX, AX, BX, and AB as having two independent packets, and the cases of A, B, and X as having one packet). For benchmarking, we first derive a theoretical upper bound for the overall MAC-layer normalized throughput imposed by the PHY-layer received data. The upper bound of NCMA is given by
\begin{align}
\text{Upper} &\text{~Bound} =2 \times (\Pr \{ ABX\} + \Pr \{ AX|BX\} \notag \\
&+ \Pr \{ AB\}) +1  \times (\Pr \{ A|B\}  + \Pr \{ X\}).
\label{equ:upper_bound}
\end{align}

We remark that, for the same normalized throughput, the absolute throughputs for QPSK and 16-QAM are twice and four times that of BPSK, respectively. The upper bound in (\ref{equ:upper_bound}) and experimental results below should be interpreted in this light.

We examine the MAC-layer performance by employing \emph{trace-driven} simulations using the PHY-layer statistics obtained in Fig. \ref{fig:phy_stat_benchmark} and Fig. \ref{fig:phy_stat_mimoncma}. We first can obtain the probabilities of all events (i.e., ABX, AB, AX$|$BX, A$|$B and X). Then, we generate traces based on the PHY-layer statistics to drive our MAC-layer simulations.

The normalized throughput for NCMA systems is defined as
\begin{align}
Th = \frac{{{L_A} \times {N_A} + {L_B} \times {N_B}}}{{{N_{Beacon}}}},
\label{equ:throughput}
\end{align}
\noindent where $N_A$ ($N_B$) is the number of messages that node $A$ ($B$) have been recovered,  ${N_{Beacon}}$ is the number of beacons and $L_A$ ($L_B$) is the number of packets the AP needs to decode $M^A$ ($M^B$)\footnote{In NCMA, the MAC-layer RS code's constraint length parameter $L$ (see Section \ref{sec:Overview2}) can be different for different nodes. We choose ${L_A} = 1.5{L_B} = 24$ based on our prior experimental results: the detailed explanation and justification for using asymmetric $L_A$ and $L_B$ can be found in \referred{NCMA1}\cite{NCMA1}.}.


Fig. \ref{fig:mac_throughput_qpsk} plots the MAC-layer normalized throughputs of different schemes using QPSK. In Fig. \ref{fig:mac_throughput_qpsk}, QPSK MIMO-NCMA's achievable throughput almost coincides with the theoretical upper bound (we show in Section \ref{sec:Exp22} that both PNC and MUD symbol-level decoders work better than their corresponding bit-level decoders, so here we only present the upper bound of NCMA using symbol-level decoders, namely the Upper Bound (Symbol) in Fig. \ref{fig:mac_throughput_qpsk} and Fig. \ref{fig:mac_throughput_16QAM}). From Fig. \ref{fig:mac_throughput_qpsk}, we can see that MIMO-NCMA works well with QPSK, and double the throughput of BPSK NCMA. We also include conventional MMSE and ZF decoders as our benchmarks in Fig. \ref{fig:mac_throughput_qpsk}. MIMO-NCMA has around 80$\sim$100\% throughput improvement over them for all SNRs.


Note that in Fig. \ref{fig:mac_throughput_qpsk}, the QPSK MAC-layer throughput for the symbol-level decoder improves very little compared with the bit-level decoder. That is, the joint use of MIMO and bit-level decoding has already provided good performance for QPSK NCMA.

In contrast, the 16-QAM MAC-layer throughput for the symbol-level decoder improves more significantly, as shown in Fig. \ref{fig:mac_throughput_16QAM}. The MAC-layer throughput result is consistent with the PHY layer decoding results, i.e., both 16-QAM symbol-level PNC and MUD decoders outperform their corresponding bit-level decoders significantly (see Fig. \ref{fig:symbol_vs_bit}(b)).  That is, for 16-QAM, both MIMO and symbol-level decoding are needed for good performance.

We also find that, as expected, 16-QAM requires higher SNRs than QPSK (e.g., to achieve a normalized throughput of one packet per time slot, 16-QAM requires 18dB while QPSK only requires 9dB). The saturated normalized throughput for 16-QAM is saturated at around 1.5 packets per time slot, which is lower than 1.8 packets for QPSK. This is because in 16-QAM, mapping overlapped constellation points (i.e., different symbol pairs) into different XOR symbols $-$ an ambiguity problem $-$ are inevitable due to the dense constellation map, and the number of constellation points in each cluster may be much larger than that in QPSK. Fortunately, with symbol-level PHY-layer decoding, the 16-QAM MIMO-NCMA can achieve 3.5 times the throughput of the previous BPSK NCMA system.

\section{Conclusions}\label{sec:Conclusions}
We have developed the first NCMA system that operates on high-order modulations beyond BPSK. In particular, we have demonstrated the feasibility of NCMA operated with QPSK and 16-QAM modulations. Moving from BPSK to these high-order modulations presents a number of challenges, among which the phase offset penalty is the most critical one. To tackle the phase penalty problem while retaining decoding simplicity (we use only non-iterative decoding), we put forth the idea of using a combination of two antennas and symbol-level decoding at the AP.  We refer to our high-order NCMA system as \emph{MIMO-NCMA}. Experiments on our software defined radio prototype show that at SNR = 10dB, our MIMO-NCMA system with QPSK modulation can double the throughput of the previous BPSK NCMA system. At SNR=20dB, the throughput of MIMO-NCMA with 16-QAM is 3.5 times that of BPSK NCMA. Overall, our results indicate that MIMO-NCMA is a promising technique to boost the throughput of NCMA at medium to high SNRs.


\appendices
\section{Reduced Complexity MUD Demodulator} \label{sec:appendix_MUD}
This appendix presents the reduced complexity NCMA demodulator (using 16-QAM as an example), and focuses on the symbol-level demodulator. In particular, we show that for MUD, the constellation of the received signal can be reshaped into a standard 16-QAM constellation \referred{dot11std13}\cite{dot11std13} after equalization on one user, and doing so reduces  demodulation complexity. We further demonstrate that the BER of the reduced complexity MUD decoder is exactly the same as that of the exhaustive-search MUD decoder studied in Section \ref{sec:Symbol-level_NCMA}. In Appendix \ref{sec:appendix_PNC}, we adapt the reduced complexity principle of MUD for PNC.

\begin{figure}[t]
\centering
\includegraphics[width=0.46\textwidth]{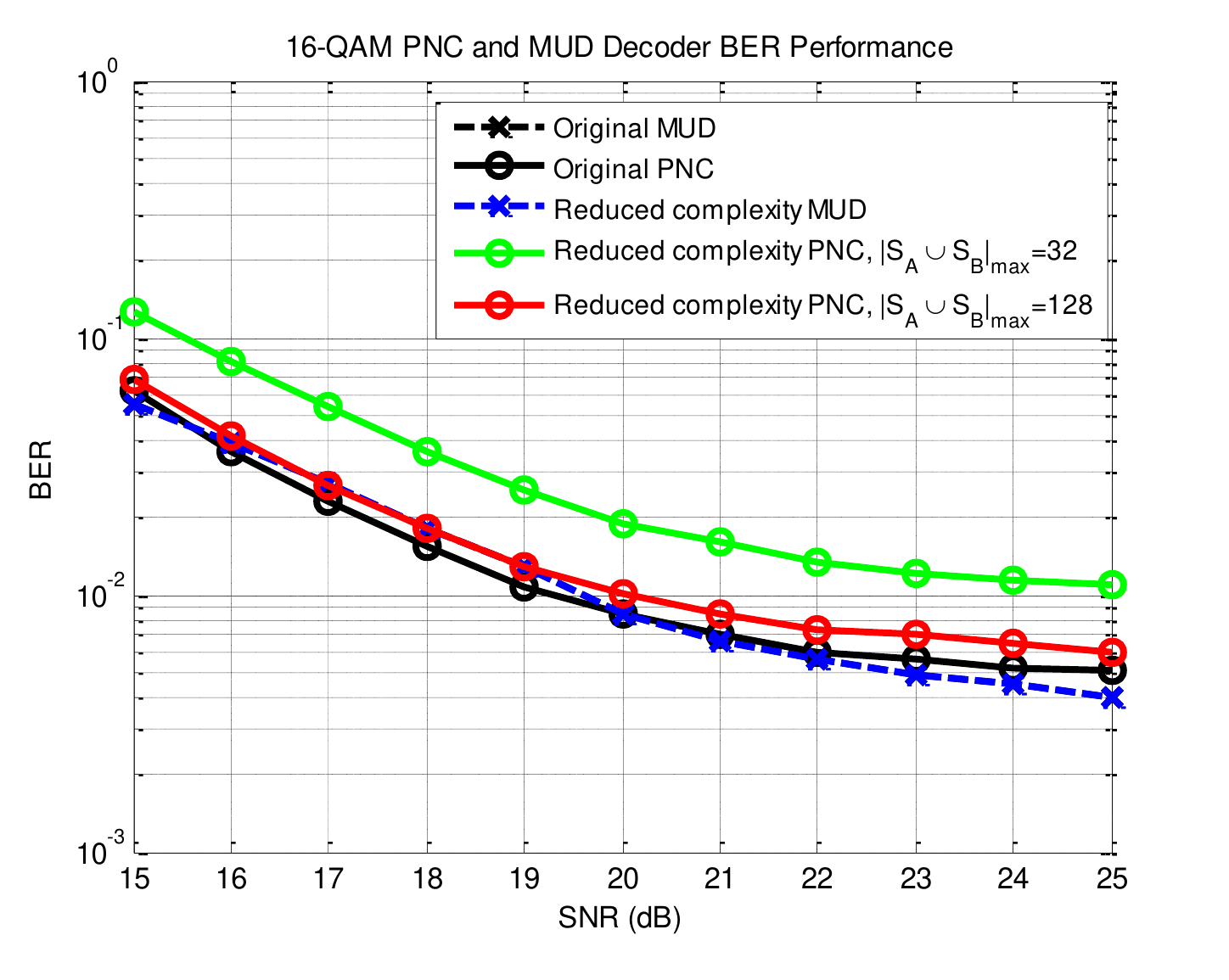}
\caption{Experimental BER performance for 16-QAM MUD and PNC decoder with reduced complexity demodulators and original demodulators. The reduced-complexity MUD decoder has the same BER performance, and the reduced-complexity PNC decoder has less than 1dB performance loss. ${S_A}$ (${S_B}$) is the set of selected points to represent  ${x_A}$ (${x_B}$), and $| \cdot {|_{\max }}$ denotes the maximum size of a set.}
\label{fig:ber_16qam_reduced}
\end{figure}

Without loss of generality, let us focus on the demodulation of node A's information. To obtain the soft information of  $x_A$, we need to find 16 representative constellation points associated with different $x_A$ out of the 256 possible input symbol pairs $(x_A,x_B)$  based on the received samples $({y_{R1}},{y_{R2}})$ obtained in (\ref{equ:y1y2}). To calculate the soft information of a particular realization of $x_A$, say ${\bar x_A}$, (\ref{equ:y1y2}) can be expressed as
\begin{align}
{y_{R1}} = {h_{A1}}{\bar x_A} + {h_{B1}}{x_B} + {w_1}, \notag \\
{y_{R2}} = {h_{A2}}{\bar x_A} + {h_{B2}}{x_B} + {w_2},
\label{equ:y1y2_appendix}
\end{align}
Since we have two equations with one unknown variable   (which can take any value of a 16-QAM symbol), maximum ratio combining (MRC) \referred{tse2005fundamentals}\cite{tse2005fundamentals} can be adopted. We have
\begin{align}
&\underbrace {h_{B1}^*{y_{R1}} + h_{B2}^*{y_{R2}} - (h_{B1}^*{h_{A1}} + h_{B2}^*{h_{A2}}){{\bar x}_A}}_{{y^{MRC}}} \notag \\
&= \underbrace {(|{h_{B1}}{|^2} + |{h_{B2}}{|^2})}_{h_B^{MRC}}{x_B} + \underbrace {h_{B1}^*{w_1} + h_{B2}^*{w_2}}_{{w^{MRC}}},
\label{equ:mrc_appendix}
\end{align}
\noindent where $*$ denotes the complex conjugate operation. Through equalization, i.e., dividing both sides of \label{equ:mrc_appendix} by $h_B^{MRC}$ , we have $\frac{{{y^{MRC}}}}{{h_B^{MRC}}} = {x_B} + \frac{{{w^{MRC}}}}{{h_B^{MRC}}}$. The estimate of ${\hat x_B}$ can therefore be demodulated using the fixed 16-QAM grid decision regions with standard approaches. The soft information of ${\bar x_A}$ (e.g., the representative Euclidean distance for the inputs of the Viterbi channel decoder) can be then expressed as
\begin{align}
d_{{{\bar x}_A}}^2 = |{y_{R1}} - {h_{A1}}{{\bar x}_A} - {h_{B1}}{{\hat x}_B}{|^2} + |{y_{R2}} - {h_{A2}}{{\bar x}_A} - {h_{B2}}{{\hat x}_B}{|^2} .
\label{equ:soft_info_appendix}
\end{align}
Instead of finding the smallest value among 16 points through distance computation, we can simply find the nearest ${\hat x_B}$ based on the fixed 16-QAM decision regions. Similarly, to obtain the soft information of $x_B$, we can find one constellation point $({\hat x_A},{\bar x_B})$ for each ${x_B} = {\bar x_B}$, and then compute the soft information $d_{{{\bar x}_B}}^2$. In Fig. \ref{fig:ber_16qam_reduced}, the BER performance based on our experimental data shows that the SIC-based reduced complexity MUD decoder has exactly the same BER performance as that of the original exhaustive-search one, but with much less demodulation time (e.g., around 10\% of the exhaustive search, see Table \ref{tab:decodingtime} in Section \ref{sec:Exp23}).

\section{Reduced Complexity PNC Demodulator} \label{sec:appendix_PNC}
For the MUD demodulator, we can make use of SIC and MRC to reduce the demodulation complexity. However, for an optimal PNC demodulator, the decision on ${x_{A \oplus B}}$ has to be made based on a joint symbol pair $({x_A},{x_B})$, rather than performing MUD marginalization first and then XOR \referred{liew2015primer}\cite{liew2015primer}. As a result, the optimal PNC demodulation region is irregular since we cannot perform equalization to get rid of two different channel gains of nodes A and B simultaneously. In this appendix, we try to make a balance between performance and complexity by extending the MUD complexity reducing scheme to approach the performance of an optimal PNC demodulator. We remark that the proposed reduced complexity PNC demodulator will depredate to the MUD demodulator presented in Appendix \ref{sec:appendix_MUD} if we only consider the 16 selected constellation points for each of ${x_A}$ and ${x_B}$ in MUD demodulation; on the other hand, the reduced complexity PNC demodulator becomes the exhaustive-search optimal PNC demodulator if we consider all the 256 constellation points. In the following, we use the MUD complexity reduction scheme as an example to explain the reduced-complexity principle for PNC demodulator.

Let ${S_A}$ and ${S_B}$ denote the sets of selected constellation points that represent ${x_A}$ and ${x_B}$ in the reduced complexity MUD demodulator, respectively (e.g., we select one nearest point to represent each ${x_A} = {\bar x_A}$  and  ${x_B} = {\bar x_B}$, such that both ${S_A}$ and ${S_B}$ contain 16 constellation points). We find representatives of ${x_{A \oplus B}}$  from ${S_A} \cup {S_B}$ (the union of sets ${S_A}$ and ${S_B}$) since each constellation point will be mapped to its corresponding XOR symbol ${x_{A \oplus B}}$. Suppose that we want to find the representative of ${x_{A \oplus B}} = {\bar x_{A \oplus B}}$. We search for all possible $({x_A},{x_B})$ that map to ${\bar x_{A \oplus B}}$ in ${S_A} \cup {S_B}$, and select the most likely one with the smallest Euclidean distance. We refer to this PNC demodulator as \emph{nearest-point demodulator}.

Fig. \ref{fig:ber_16qam_reduced} shows that with the nearest-point demodulator (i.e., the green curve), the reduced complexity PNC decoder has around 2dB BER performance loss compared with the optimal PNC decoder (i.e., the black curve). The reason for the BER performance degradation is that we only consider a small subset of the total 256 constellation points (i.e., the maximum size of ${S_A} \cup {S_B}$ is 32), and it is difficult to find all the 16 XOR representatives for most of the time.

To narrow down the performance gap, we need to consider more constellation points. For example, we can select more than one nearest point to represent each ${x_A} = {\bar x_A}$ (${x_B} = {\bar x_B}$). Let us consider four nearest points, which follows the sphere decoding principle \referred{tse2005fundamentals}\cite{tse2005fundamentals}. The size of ${S_A}$ or ${S_B}$ now becomes 64, and the maximum size of ${S_A} \cup {S_B}$ is 128 (the union set is smaller than 128 if there are same pairs in ${S_A}$ and ${S_B}$). The resultant BER performance can be improved (i.e., the red curve in Fig. \ref{fig:ber_16qam_reduced}), which almost coincides with the original exhaustive-search PNC decoder. For PNC decoder, there exists a fundamental demodulation complexity and BER performance tradeoff.

\begin{table*}
\caption{\textnormal{Complexity analysis and decoding times of bit-level and symbol-level channel decoders: (a) Theoretical complexity with M-QAM, where $N$ is the number of source bits, $K$ is the number of states and $r$ is the coding rate ($K$=64, $r$=1/2 in this paper). For M-QAM's trellis in symbol-level decoder ($M \geq 4$), each state has ${2^{r{{\log }_2}M}}$ output branches, and the number of stages reduces to $N/(r{\log _2}M)$. (b) Channel decoding times (for 16-QAM) of SPIRAL and GNU Radio decoders.} }
\vspace*{-10pt}
\centering
\begin{subtable}{0.47\linewidth}
\caption{}
\begin{tabular}{|c|c|c|} \hline
\backslashbox {\scriptsize{Modulation}}{\scriptsize{Decoder}}    &    Bit-level             & Symbol-level            \\ \hline \hline
BPSK  &  2KN      &   N.~A.    \\ \hline
QPSK   &  2KN      & 2KN   \\ \hline
16-QAM  &  2KN      & 2KN       \\ \hline
64-QAM   &  2KN      & 8KN/3   \\ \hline
M-QAM    &  2KN   & ${2^{r{{\log }_2}M}} \cdot K \cdot N/(r{\log _2}M)$ \\ \hline
\end{tabular}
\end{subtable}
\qquad
\begin{subtable}{0.48\linewidth}
\caption{}
\begin{tabular}{|c|c|c|c|c|} \hline
        \multirow{2}{1.95cm}{Spiral Optimized  (ms) \referred{SpiralSOVA}\cite{SpiralSOVA}}
                & \multirow{2}{1.8cm}{Spiral Standard  (ms) \referred{SpiralSOVA}\cite{SpiralSOVA}}  &
        \multirow{2}{1.65cm}{GNU Radio Bit-level (ms)} & \multirow{2}{2.15cm}{GNU Radio Symbol-level (ms)} \\
                ~& ~& ~ & ~ \\  \hline  \hline
                 0.48 & 2.21 & 2.18 & 2.29\\
        \hline
\end{tabular}
\end{subtable}
\label{tab:complexity}
\end{table*}

\begin{table*}[t]
\centering
\caption{\textnormal{Demodulation time comparison between bit-level and symbol-level NCMA demodulators using GNU Radio.}}
\begin{tabular}{|c|c|c|c|c|c|l|c|c|}
 \hline
 \backslashbox {\scriptsize{Modulation}}{\scriptsize{Demodulator}} &
 \multirow{2}{2.2cm}{Point-to-point (ms)} &
 \multicolumn{3}{c|}{Bit-level (ms)} &
 \multicolumn{3}{c|}{Symbol-level (ms)}  \\
 \cline{3-8}
   & & PNC & MUD & NCMA  & PNC & MUD & NCMA\\
 \hline \hline
 BPSK & 0.29 & 0.72 & 0.79 & 0.81 & N.A. & N.A. & N.A. \\  \hline
 QPSK & 0.79 &1.72 &1.92 &2.23 &1.84 &1.96 &2.40\\  \hline
 16-QAM & 1.99 &113.54 &124.63 &137.09 &149.96 &160.71 &180.36 \\  \hline
 16-QAM   & & & & & & & \\
 (reduced complexity) & N.A. &58.46 &16.49 &58.87 &60.72 &16.51 &60.92 \\ \hline
 \end{tabular}
\label{tab:decodingtime}
\end{table*}

\section{Performance Improvement Comparison between Symbol-level PNC and MUD Decoders} \label{sec:appendix_reason}
In the simulation and experimental results, we find that symbol-level decoding brings substantial improvement over bit-level decoding for the PNC decoder. By contrast, symbol-level decoding brings smaller improvement over bit-level decoding for the MUD decoder. In this appendix, we give an intuitive explanation for this phenomenon using QPSK as an example (the analysis for 16-QAM follows similar methods).

To understand the performance improvement of the symbol-level PNC decoder, let us look at the four constellation points overlapping at the origin of Fig. \ref{fig:Constellation}. The symbol-level demodulator maps (demodulates) the four overlapping constellation points $(1+j,-1+j)$, $(-1-j,1-j)$, $(1-j,1+j)$ and $(-1+j,-1-j)$ to two possible QPSK XOR symbols $-1+j$ and $1-j$ (i.e., we get two possible XORed symbol pairs: (-1, 1) and (1, -1)). And these two QPSK XOR symbols should have the same probability as the soft information for the XOR symbol, considering all the four possible input pairs are overlapped. This is useful information for the symbol-level decoder because the decoder can rule out two other XOR combinations: (1, 1) and (-1, -1). The PNC bit-level demodulator, on the other hand, will map the four constellation points to four XOR pairs: (-1,-1), (-1, 1), (1, -1) and (1, 1). Hence, no useful information about the input bits can be obtained through the bit-level demodulation process (since each XOR bit is equally likely to be 1 or -1). In this example, the number of overlapped constellation points can be reduced to a smaller number of XOR symbols through PNC mapping, and the symbol-level PNC demodulator can avoid huge information loss. Therefore, we observe a large improvement of the symbol-level QPSK PNC decoder over the bit-level one (e.g., see Fig. \ref{fig:symbol_vs_bit}(a)).

However, with the symbol-level MUD demodulator, the number of possible demodulated MUD symbols always remains the same as the number of the overlapped constellation points. Therefore, the symbol-level demodulator may still have the same results as the bit-level one. To see an intuitive example, let us use the four overlapped constellation points at the origin again. There are four possible symbols for node A (namely, $1+j$, $-1-j$, $1-j$ and $-1-j$) with equal probability when we marginalize user B's information. Thus, the symbol-level MUD decoder also has no useful information after demodulation, and the improvement of the symbol-level MUD decoder is not as significant as the PNC decoder. For QPSK, the symbol-level MUD decoder only improves the performance of bit-level MUD slightly (e.g., see Fig. \ref{fig:symbol_vs_bit}(a), and the small improvement comes from overlapping constellation points with fewer potential input combination possibilities).

\section{Decoding Time Measurements for PHY-layer Decoders} \label{sec:appendix_complexity}
In the appendix, we present decoding time measurements for the PHY-layer MIMO-NCMA decoders, including the two main components: (1) the channel decoder, and (2) the demodulator.

We perform experiments to evaluate the real-time decoding speeds of MIMO-NCMA PHY-layer decoders. The implementation of the symbol-level Viterbi decoder was developed using GNU Radio \emph{gr-trellis} library \referred{GNURadio}\cite{GNURadio}. The average processing times (the mean value of 10,000 packets), for channel decoder and demodulator, and shown in Table \ref{tab:complexity}(b) and Table \ref{tab:decodingtime}, respectively.

For channel decoders, we use 16-QAM modulation to test their speed performance. As a benchmark, in Table \ref{tab:complexity}(b), we compare the channel decoding times (excluding the demodulation time) with the bit-level SPIRAL Viterbi decoder that has been optimized at the machine code level \referred{SpiralSOVA}\cite{SpiralSOVA}. We can see that there is a big gap (i.e., 5 times) between the GNU Radio decoder and the SPIRAL optimized decoder. However, the standard SPIRAL decoder and GNU Radio are comparable in speed. Furthermore, the symbol-level and bit-level GNU Radio decoders have comparable speed. The experimental processing time corroborates the theoretical complexity analysis in Table \ref{tab:complexity}(a). We believe that the symbol-level channel decoding latencies can be further reduced following similar approaches like SPIRAL.

For demodulators, we can see from Table \ref{tab:decodingtime} that the processing time for QPSK MIMO-NCMA is small and comparable to the previous BPSK single-antenna case. However, the demodulation time for 16-QAM increases significantly compared with that of QPSK, because both 16-QAM bit-level and symbol-level demodulators have to compute 256 Euclidean distances, while QPSK only needs to compute 16 Euclidean distances. Recall that, in Secion \ref{sec:Symbol-level_NCMA3} and Appendix \ref{sec:appendix_MUD}, we find that MUD demodulator's complexity can be reduced without sacrificing any BER performance. The decoding times for our proposed reduce-complexity demodulators are also shown in Table \ref{tab:decodingtime}. We can see that the demodulation time is greatly reduced after simplification. 

\section*{Acknowledgment}
This work is supported by AoE grant E-02/08 and the General Research Funds Project Number 14204714 and 14226616, established under the University Grant Committee of the Hong Kong Special Administrative Region, China. This work is also supported by the China NSFC grants (Project No. 61271277 and No. 61501390)

\bibliographystyle{IEEEtran}
\bibliography{mimo-ncma}

\end{document}